\documentclass[11pt,a4paper]{article}
\pdfoutput=1
\usepackage{jcappub}
\usepackage{graphicx}
\usepackage{epsfig}
\usepackage{amsmath, amssymb}
\usepackage{dcolumn}
\usepackage{bm}
\usepackage{amsfonts}
\usepackage[toc,page]{appendix} 
\usepackage{color}

\usepackage[flushleft]{threeparttable}

\usepackage{subfigure}
\usepackage{amssymb}
\usepackage{epstopdf}

\newcommand{\beq}{\begin{equation}}
\newcommand{\eeq}{\end{equation}}
\newcommand{\bea}{\begin{eqnarray}}
\newcommand{\eea}{\end{eqnarray}}
\newcommand{\ba}{\begin{array}}
\newcommand{\ea}{\end{array}}
\newcommand{\bit}{\begin{itemize}}
\newcommand{\eit}{\end{itemize}}

\newcommand{\bi}{\begin{itemize}}
\newcommand{\ei}{\end{itemize}}

\newcommand{\ie}{\textit{i.e.\ }}
\newcommand{\eg}{\textit{e.g.\ }}

\newcommand{\xiL}{\xi_{\mathcal{L}}}
\newcommand{\xiW}{\xi_{\rm W}}
\newcommand{\fp}{f_{\rm pq}^{\rm p}}
\newcommand{\ftot}{f_{\rm pq}^{\rm Tot}}
\newcommand{\lpq}{\ell_{\rm pq}}
\newcommand{\vf}{\langle {\rm v}^2\rangle_{\rm F}}
\newcommand{\vg}{\langle {\rm v}^2\rangle_{\rm G}}

\newcommand{\vpq}{\langle {\rm v}^2\rangle^{\rm pq}}
\newcommand{\vpqf}{\langle {\rm v}^2\rangle_{\rm F}^{\rm pq}}
\newcommand{\vpqg}{\langle {\rm v}^2\rangle_{\rm G}^{\rm pq}}

\begin{document}

\title{Survival of $pq$-superstrings in field theory simulations}

\author[]{Joanes Lizarraga, }
\author[]{Jon Urrestilla}
\affiliation[]{Department of Theoretical Physics, University of the Basque Country UPV/EHU, 48040 Bilbao, Spain}
\emailAdd{joanes.lizarraga@ehu.eus}
\emailAdd{jon.urrestilla@ehu.eus}

\date{\today}

\abstract{We perform large-scale field theoretical simulations in expanding universe to characterize a network of strings that can form composed bound states. The network consists of two copies of Abelian Higgs strings (which we label $p$ and $q$, respectively) coupled via a potential term to give $pq$  bound states. The simulations are performed using two different kinds of initial conditions: the first one  with a network of $p$- and $q$-strings, and the second one with a  network of $q$- and $pq$-strings. This way, we start from two opposite situations: one with no initial $pq$-strings, and one with a large initial number of $pq$-strings.  We find that in both cases the system scales, and in both cases the system prefers to have a low fraction of $pq$-strings. This is somewhat surprising in the case for the second type of conditions, showing that the unzipping mechanism is very efficient. We also find hints that both initial conditions tend to asymptote to a common configuration, though we would need a larger dynamical range to confirm it. The average velocities of the different types of strings in the network have also been explored for the first time.}

\maketitle


\section{Introduction}
\label{sec_intro}

The inflationary scenario has successfully been supported by modern and accurate observations \cite{Dvorkin:2011aj,Hinshaw:2012aka,Ade:2013xla,Ade:2013zuv} as the best model for the origin of density fluctuations and the observed large-scale structure formation. However, the fundamental physics behind inflation still remains as one of the main unanswered questions of modern cosmology. Brane inflation models, proposed in the context of superstring theories, are interesting candidates to provide such an answer. It is expected that fundamental strings as well as string-like objects, such as $D$-branes, could form at the end of brane inflation due to brane-antibrane annihilation processes \cite{Majumdar:2002hy,Sarangi:2002yt,Jones:2003da,Dvali:2003zj,Copeland:2003bj}. Such \textit{superstrings} could survive the cosmological evolution and can have cosmological size \cite{Polchinski:2004ia,Davis:2005dd}, which implies that they can play a role similar to ordinary cosmic strings. Therefore, their possible observational signal could be measurable, which would be invaluable towards an experimental verification of string theories and a remarkable window into the physics behind inflation.

Superstring networks produced after brane annihilation processes differ from ordinary cosmic string networks. Unlike solitonic strings, cosmic superstrings do not always intercommute; for example, they can join together forming heavy and stable bound states composed of $p$ $F$-strings and $q$ $D$-strings called $pq$-strings \cite{Dvali:2003zj,Copeland:2003bj}. The boundaries of these segments are determined by 3-way Y-junctions where 3 different strings meet. Y-junctions are particularly interesting because they can perturb and modify the expected evolution of the network, producing observable effects or even preventing the network from achieving the scaling regime \cite{Sakellariadou:2004wq}. 

The evolution and description of $F$ and $D$ interconnected networks is rather complicated and numerical explorations are highly desirable. Early works on cosmic superstring networks analyze non-Abelian field theories of the kind ${\rm G} \rightarrow {\rm Z}_{3}$, where 3-way type junctions can form, both analytically \cite{Aryal:1986cp,Vachaspati:1986cc} and numerically \cite{Spergel:1996ai,Hindmarsh:2006qn}. More sophisticated analytical models include effective field theories such as the velocity-dependent one-scale (VOS) models (see for instance \cite{Tye:2005fn,Avgoustidis:2007aa}) where different type of strings with different tension ranges can be considered. However, there is no consensus among those works on the physics behind the energy loss mechanism that leads to scaling of $FD$-networks, specifically whether the excess energy is radiated or, in turn, goes to increase the kinetical energy of the strings.

The evolution of $F$ and $D$-strings can also be modeled by field theoretical {\it ordinary} solitonic strings, which are formed by usual spontaneous symmetry breaking mechanisms \cite{Saffin:2005cs,Rajantie:2007hp}. These models consider a system composed by a pair of complex scalar fields, which is invariant under $\mathrm{U}(1)\times \mathrm{U}(1)$. The formation of stable bound states is typically accomplished by modifying the form of the potential of the system.

Full field theory simulations have been employed to explore the validity of these field theoretical models in reproducing the interconnected string dynamics \cite{Urrestilla:2007yw,Rajantie:2007hp,Sakellariadou:2008ay}. Using the previously mentioned models, field theory simulations demonstrated that interconnected string networks can form at spontaneous symmetry breaking phase transitions. The main goal of the works \cite{Rajantie:2007hp,Sakellariadou:2008ay} was to measure the effects produced by long range interactions in the overall dynamics of the network and bound states. Long range interactions were included in the analysis considering that one of the two ${\rm U}(1)$ symmetries of the system was global. They also performed simulations were an {\it already-formed} network of $pq$-strings was present from the beginning. They found that, regardless of the initial conditions, the relative amount and significance of the bound states was considerably low, \ie long range effects mediated by massless Goldstone bosons tend to break bound states. Conversely, they observed that in absence of long range interactions, strings of the initial $pq$-network remain in bound states for much longer. Scaling of such networks was also confirmed in \cite{Sakellariadou:2008ay}.

Field theoretical simulations of the model presented in \cite{Saffin:2005cs}, where both ${\rm U}(1)$ are local, were analyzed in \cite{Urrestilla:2007yw}. In this work only short range interactions  were studied, since long range interactions other than gravity were expected to be of little relevance in $FD$-networks. The authors successfully confirmed that interconnected string networks modeled by this model reach the scaling regime, which is an indispensable requisite for the cosmological viability of defects. However, as in previously mentioned works, it was observed that bound states constitute only a small fraction of the total string length of the system, of about $\sim 2\%$. Moreover, it was found that the length and lifetime of the bound states were very short.

All these early works, thus, put forward an interesting debate regarding the amount, lifetime and relevance of $pq$-strings. The observed formation rate and abundance of heavy strings are low, much lower than predicted theoretically by analytical models \cite{Tye:2005fn,Avgoustidis:2007aa}, where they show that the number density of $pq$-strings is comparable to the number densities of $p$- and $q$-strings. This discrepancy has recently been linked to the role played by Y-junctions in the networks, whose importance in the formation and shrinking of bound states may be relevant. Motivated by the tendency exhibited by field theory simulations, in \cite{Avgoustidis:2014rqa} the stabilization and unzipping process conditions have been explored. Unzipping of heavy strings might be an extra ingredient to take into account in the development of effective models.

The main goal of this work is to extend previous field theoretical simulations performing the biggest and most accurate field theory simulations of cosmic superstrings. In  \cite{Urrestilla:2007yw}, the network of strings was found to scale, and form bound states, although a very low fraction of the total length of string was in bound states.  One question raised in that work was whether the limited dynamical range of the simulation was the reason for the low fraction, since a big part of the simulation time went into first forming the network, and then slowly populating it with $pq$-strings. In order to give a more detailed insight on the late time evolution and relative amount of bound states, we investigate a set of simulations called {\it combined simulations}, which incorporate an artificial whole network of bound states coexisting with a network of single cosmic strings from the very beginning, similar to the initial conditions used in \cite{Rajantie:2007hp}. This perspective provides a wider view of the decay or unzipping of bound states as well as the interaction with \textit{individual} strings. The average velocity distributions of the network and $pq$-segments have also been explored for the first time, which, together with the unzipping mechanism, could be fundamental in order to build proper effective theories.

This paper is structured in the following way: in Sec.~\ref{sec_modelandimplementation} we review the model and explain the numerical procedures utilized in this work to perform the numerical simulations and identify $pq$-segments. After that, in Sec.~\ref{sec_combine} we introduce the new string combination procedure and present the results in Sec.~\ref{sec_results}. Finally we discuss the results in Sec~\ref{sec_discussion}.


\section{Description of the model and the numerical setup}
\label{sec_modelandimplementation}

\subsection{Model and parameter choice}
\label{subsec_model}

The model considered in this work was proposed in \cite{Saffin:2005cs} and posseses a $\mathrm{U}(1)_{\rm L} \times \mathrm{U}(1)_{\rm L}$ gauge symmetry, which leads to a pair of independent local cosmic strings. In addition to the usual \textit{mexican hat} potential for the symmetry breaking, doubled in this case, the potential of this model also contains an extra interaction term which leads to the formation of stable bound states where both complex fields wind simultaneously. The whole Lagrangian reads,

\begin{equation}
\mathcal{L} = (D_{\mu}\phi)^* (D^{\mu}\phi) + (\mathcal{D}_{\mu}\psi)^*(\mathcal{D}^{\mu}\psi) - \frac{1}{4e^2}F_{\mu\nu}F^{\mu\nu} - \frac{1}{4g^2}\mathcal{F}_{\mu\nu}\mathcal{F}^{\mu\nu} - V(|\phi|,|\psi|)\, ,
\label{eq_lag}
\end{equation}
where $\phi$ and $\psi$ are the two complex scalar fields, which represent each sector of the double ${\rm U}(1)$ gauge symmetry. Each of them is independently charged with respect to its ${\rm U}(1)$ gauge field: $\phi$ is coupled to the gauge field $A_{\mu}$ with coupling constant $e$, and $\psi$ is coupled to $B_{\mu}$ with  coupling constant $g$. The covariant derivatives and the field strength tensors are then,

\begin{eqnarray}
D_{\mu}\phi = \partial_{\mu}\phi - iA_{\mu}\, , \\
\label{eq_model1}
F_{\mu\nu} = \partial_{\mu}A_{\nu} - \partial_{\nu}A_{\mu}\, , \\
\mathcal{D}_{\mu}\psi = \partial_{\mu}\psi - iB_{\mu}\, , \\
\mathcal{F}_{\mu\nu} = \partial_{\mu}B_{\nu} - \partial_{\nu}B_{\mu}\, .
\label{eq_model2}
\end{eqnarray}

The potential takes into account the symmetry breaking of each sector as well as the interaction between the two sectors:

\begin{equation}
V(|\phi|, |\psi|) = \frac{\lambda_{\rm A}}{4}(|\phi|^2 - \eta_{\rm A}^2)^2 + \frac{\lambda_{\rm B}}{4}(|\psi|^2 - \eta_{\rm B}^2)^2 - \kappa(|\phi|^2 - \eta_{\rm A}^2)(|\psi|^2 - \eta_{\rm B}^2)\, ,
\label{eq_potential}
\end{equation}
where $\lambda_{\rm A}$, $\lambda_{\rm B}$ and $\kappa$ are dimensionless coupling constants and $\eta_{\rm A}$ and $\eta_{\rm B}$ the vacuum expectation values for each kind of scalar fields.

Unless the last interaction term is considered, this potential describes the evolution of two independent networks of local cosmic strings. The last term includes the interaction between both type of fields, enabling the formation of stable bound states. As it can be seen from its form, it is only relevant where both fields are zero simultaneously. However, the exact nature of the critical points of the potential, \ie whether they are minima (stable) or maxima (non-stable), depends strongly on the value of $\kappa$; it is only in certain values of the parameter $\kappa$ where stable bound states are formed. As pointed out in \cite{Saffin:2005cs} stable bound segments appear only if $\kappa$ obeys the following relation:

\begin{equation}
0 < \kappa < \frac{1}{2}\sqrt{\lambda_{\rm A}\lambda_{\rm B}}.
\label{eq_klimit}
\end{equation}
Under the same conditions, the existence of gravitating bound states has also been demonstrated \cite{Hartmann:2008fi}.

Field equations of motion are derived in the temporal gauge ($A_0=B_0=0$) and evolved in a spatially flat Friedmann-Lema\^\i tre-Robertson-Walker (FLRW) background:

\begin{equation}
\ddot{\phi} + 2 \frac{\dot{a}}{a} \dot{\phi} - D_jD_j\phi = -a^2\left(\frac{\lambda_{\rm A}}{2}(|\phi|^2 - \eta_{\rm A}^2)-\kappa(\psi^2-\eta_{\rm B}^2)\right)\phi\,,
\label{eq_eom_1}
\end{equation}
\begin{equation}
\ddot{\psi} + 2 \frac{\dot{a}}{a} \dot{\psi} - \mathcal{D}_j\mathcal{D}_j\psi = -a^2\left(\frac{\lambda_{\rm B}}{2}(|\psi|^2 - \eta_{\rm B}^2)-\kappa(\phi^2-\eta_{\rm A}^2)\right)\psi\,,
\end{equation}
\begin{equation}
\dot{F}_{0j} - \partial_iF_{ij} = -2a^2e^2 \rm Im [\phi^*D_j\phi]\,,
\end{equation}
\begin{equation}
\dot{\mathcal{F}}_{0j} - \partial_i\mathcal{F}_{ij} = -2a^2g^2 \rm Im [\psi^*\mathcal{D}_j\psi]\,,
\label{eq_eom_2}
\end{equation}
\begin{equation}
-\partial_iF_{0i} = -2a^2e^2 {\rm Im }[\phi^*\dot{\phi}]\,, \quad -\partial_i\mathcal{F}_{0i} = -2a^2g^2 {\rm Im} [\psi^*\dot{\psi}]\,.
\label{eq_gauss}
\end{equation}

Here,  $a$ is the scale factor of the expanding universe, the dot derivatives represent derivatives with respect to the conformal time and the spatial derivatives are taken  with respect to the comoving coordinates. The last couple of equations are Gauss's law for each complex scalar field and rather than equations of motion they are constraints of the system. Note that the Eq.~({\ref{eq_lag}) is totally symmetric on both scalar and gauge fields, \ie $\phi \leftrightarrow \psi$ and $A_{\mu} \leftrightarrow B_{\mu}$; so are the equations of motion.


\subsection{Simulation details}
\label{sub:simdet}

In this section we give explicit details of the simulation setups. We discretize the field equations of motion (Eqs.~(\ref{eq_eom_1})-(\ref{eq_eom_2})) on a lattice using the standard lattice link variable approach \cite{Moriarty:1988fx,Bevis:2006mj}. We perform simulations on radiation and matter dominated FRLW background cosmologies. One of the most important challenges of such simulations in expanding universes is to resolve the string core and the expansion of the universe simultaneously. As the universe expands, the physical distance between adjacent points of the lattice increases, but the physical string width remains constant. Thus, strings will eventually shrink between lattice points and we will not be able to track them. One of the most used approaches to avoid such an undesirable situation is to consider time varying coupling constants:

\begin{equation}
\lambda_{\rm A} = \frac{\lambda_{{\rm A}0}}{a^{2(1-s)}}\, , \quad e=\frac{e_{0}}{a^{1-s}}\, ,
\end{equation}
and equivalently for $\lambda_{\rm B}$ and $g$. In this model, since the potential contains an extra interaction coupling constant $\kappa$, it has to also be made time dependent,

\beq
\kappa = \frac{\kappa_{0}}{a^{2(1-s)}}\, .
\eeq

This procedure is also known as the Press-Ryden-Spergel (or \textit{fat-string}) approach \cite{Press:1989id,Press:1989yh,Moore:2001px}. The parameter $s$ governs the relative width of the string: If $s=1$ we recover the original equations and the equations of motion give the true dynamics of the system. In the other extreme, if $s=0$ the physical width of the string grows with the expansion of the universe, or in other words, we obtain strings with constant comoving width. The modified equations of motion are obtained including the time dependent coupling constants into the gauge-invariant action and varying it with respect to the fields, which assures self-consistency of the equations \cite{Bevis:2006mj,Bevis:2010gj}. They read as:

\begin{equation}
\ddot{\phi} + 2 \frac{\dot{a}}{a} \dot{\phi} - D_jD_j\phi = -a^{2s}\left(\frac{\lambda_{{\rm A}0}}{2}(|\phi|^2 - \eta_{\rm A}^2)-\kappa_0(\psi^2-\eta_{\rm B}^2)\right)\phi\,,
\end{equation} 
\begin{equation}
\dot{F}_{0j} + 2(1-s)\frac{\dot{a}}{a}F_{0j} - \partial_iF_{ij} = -2a^{2s}e_0^2 \rm Im [\phi^*D_j\phi]\,,
\end{equation}
and the same for $\psi$ and $\mathcal{F}_{\mu\nu}$. 

Several previous works on field theoretical simulations showed the acceptability of $s=0$ to describe the dynamics of the system  of this approximation \cite{Bevis:2006mj,Bevis:2010gj}. Based on this, we choose the extreme case $s=0$ for this work so as to extend as much as possible the scaling period.  The statistical properties of a network using the {\it true} $s=1$ equations of motion are very similar to those of $s=0$, the differences will mostly lie inside the statistical and systematic errors. Bearing that in mind,  we choose to solve approximate equations of motion in order to gain  dynamical range. Moreover, so as  to balance between dynamical range and precision of previous works, we have increased the size of the simulation box to $1024^3$ and decreased the comoving spatial separation to $dx=0.5$, with time steps of $dt=0.1$, which had been shown to yield good resolution. Simulations were parallelized through the publicly available LatField2 library for parallel field theory simulations \cite{David:2015eya} and performed at the COSMOS Consortium supercomputer and i2Basque academic network computing infrastructure.

Our initial conditions set the modulo of the complex scalar fields at their corresponding vacuum expectation value and distribute randomly the complex scalar phases along the simulation box. Gauge fields and canonical conjugates of the scalar as well as the vector fields are set to zero. This choice of initial conditions yields a very energetic configuration, mainly due to high gradient contributions. In order to smooth this initial situation we applied a combined diffusive-dissipative phase for the first quarter of the simulation, specifically for a time period of $\Delta\tau = 64$. The first diffusive period (between $\tau_{\rm start}=50$ and $\tau=60$) consists of a cooling of the equations of motion where we use {\it fake} time-steps of $1/30$ rather than $dt=0.1$. The dissipative period, in turn, is performed by applying a non-physical damping of $0.4$ between the end of the diffusive phase and the beginning of the natural (or core growth CG)  evolution of the system (between $\tau=60$ and $\tau_{\rm CG}=114$). After $t_{\rm CG}=114$ the system has been evolved following the usual equations of motion for $s=0$ until $\tau_{\rm end}=303$.

Throughout this work, we restrict ourselves to the Bogomol'nyi limit where the parameters are related in the following way:

\begin{equation}
\lambda_{{\rm A}0}=2e_0^2\, , \quad \lambda_{{\rm B}0} = 2g_0^2.
\end{equation}

We also treat both type of strings identically, hence we set $e=g$ and $\eta_A=\eta_B$. Moreover, the typical parameter rescaling procedure is adopted and the parameters are reduced to:

\begin{eqnarray}
e_0=g_0=\frac{1}{2}\lambda_{{\rm A}0}=\frac{1}{2}\lambda_{{\rm B}0}=1\, , \\
\eta_{\rm A} = \eta_{\rm B} = 1\, .
\end{eqnarray}


\subsection{Output treatment: $pq$-segment identification}
\label{subsec_segmiden}
Local cosmic strings on a lattice can be identified in several ways, \eg using the energy density, using the value of the potential or using the windings of the phase of the field. In this work, we use the latter procedure and localize strings calculating the winding of the complex phase of each field in every lattice placket. In order to accomplish that, we use the gauge invariant definition of the winding presented in \cite{Kajantie:1998bg}. A snapshot of a $1024^3$ simulation is shown in Fig.~\ref{fig:loops}. Blue and green lines represent \textit{raw} individual $p$- and $q$-strings respectively, while red points are the points of the lattice where both fields wind simultaneously.

\begin{figure}[h!]
\centering
\includegraphics[width=0.49\textwidth]{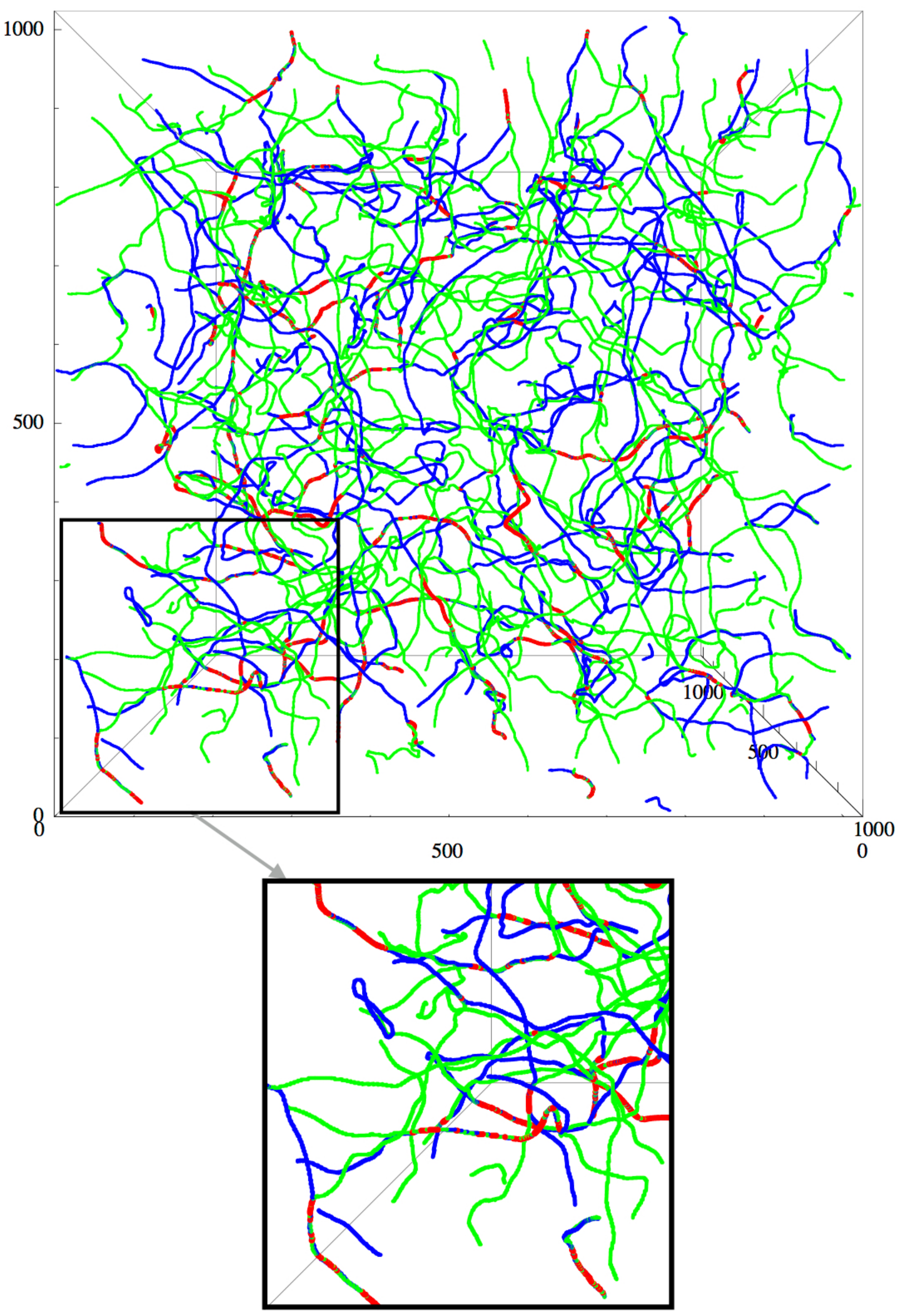}
\includegraphics[width=0.49\textwidth]{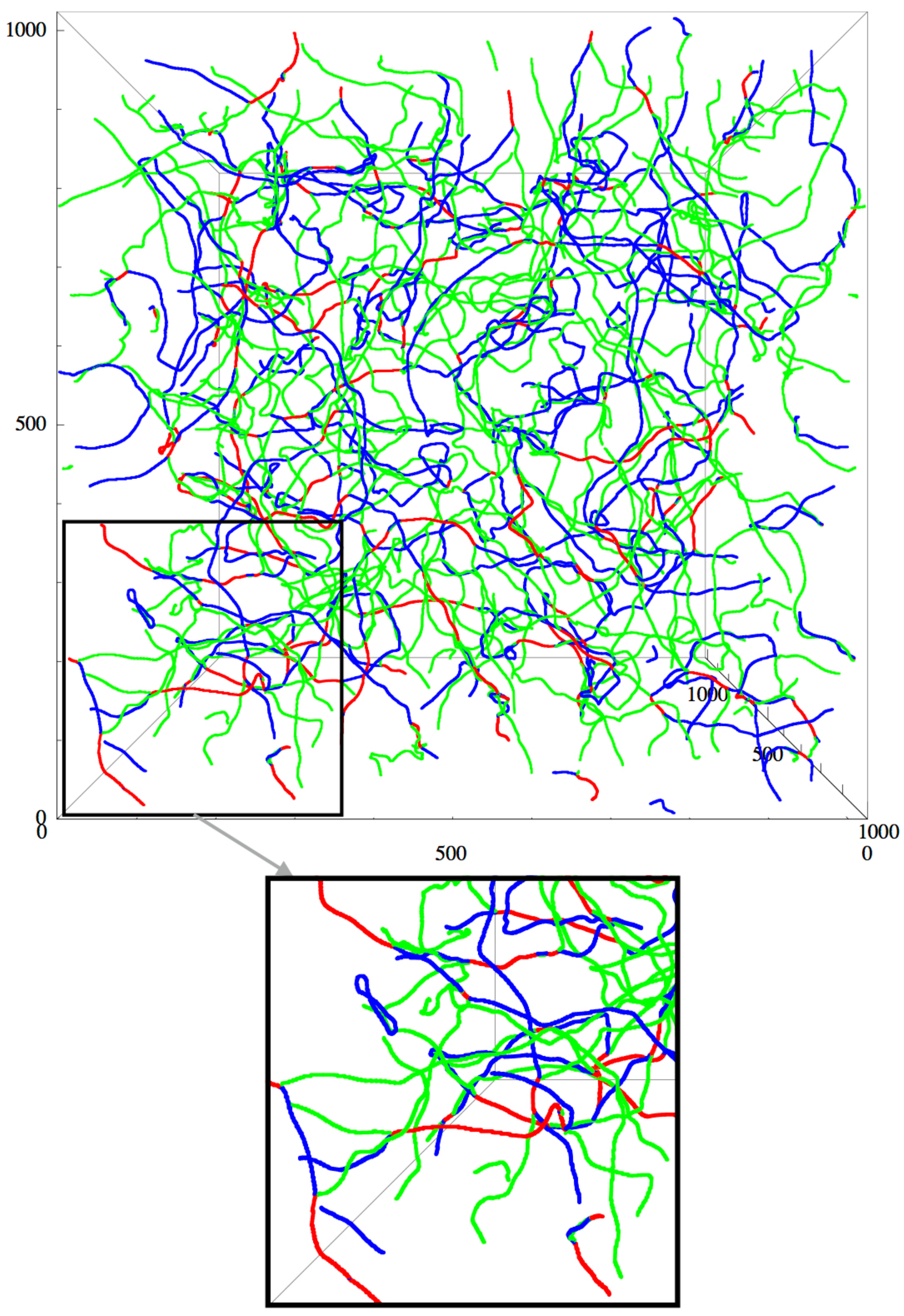}
\caption{Windings obtained from a $1024^3$ simulation. Left panel shows \textit{raw} windings where  blue and green lines represent windings of the $\phi$ and $\psi$ fields respectively. The red points represent double winding points, {\it i.e.}, points where both $\phi$ and $\psi$ wind. The right panel shows the same snapshot after the segment identification treatment explained in Sec.~\ref{subsec_segmiden} has been performed. The two lower {\it blow-up} squares show clearly how our identification procedure corrects for discontinuities in the $pq$-strings.}
\label{fig:loops}
\end{figure}

Essentially $pq$-strings are created as a consequence of the attraction produced by the interaction term in the potential and, in theory, they are located at points where both complex scalar fields wind simultaneously. However, due to the finite resolution of our simulations there are cases where although individual $p$- and $q$-string cores overlap, the winding is not located exactly at the same plaquette and might be displaced by $dx$ distance within a individual $pq$-segment, giving the impression that they do not form a bound state. In order to avoid this confusion, we set a new criterium to identify $pq$-segments: in addition to points where we find exactly double windings, segments will also be composed by regions where $p$- and $q$-strings are separated by less than 2 physical width units or in our lattice units by less than $d_{\mathrm{AB}}=4$, regardless of whether they contain exact doubly-winded points. This approach is somewhat different of that proposed in \cite{Urrestilla:2007yw}, where segments are determined considering the intersegment distance, \ie \textit{gap} distance between segments, rather than considering the transverse distance between $p$- and $q$-strings. 

\begin{figure}[h!]
\centering
\includegraphics[width=0.49\textwidth]{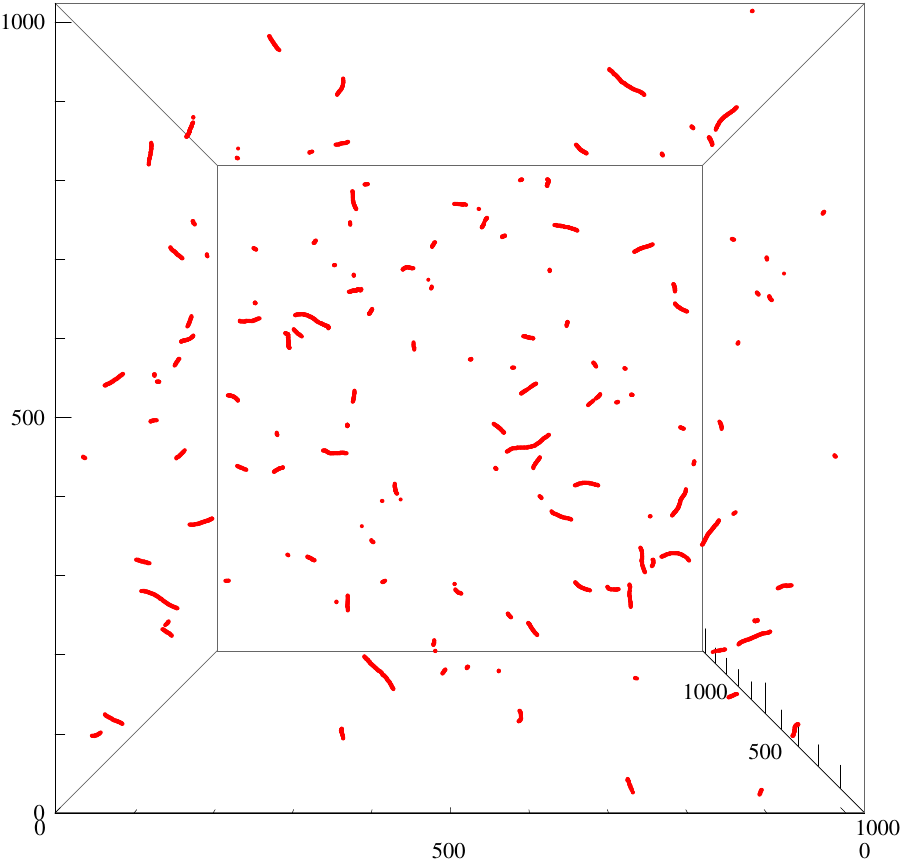}
\includegraphics[width=0.49\textwidth]{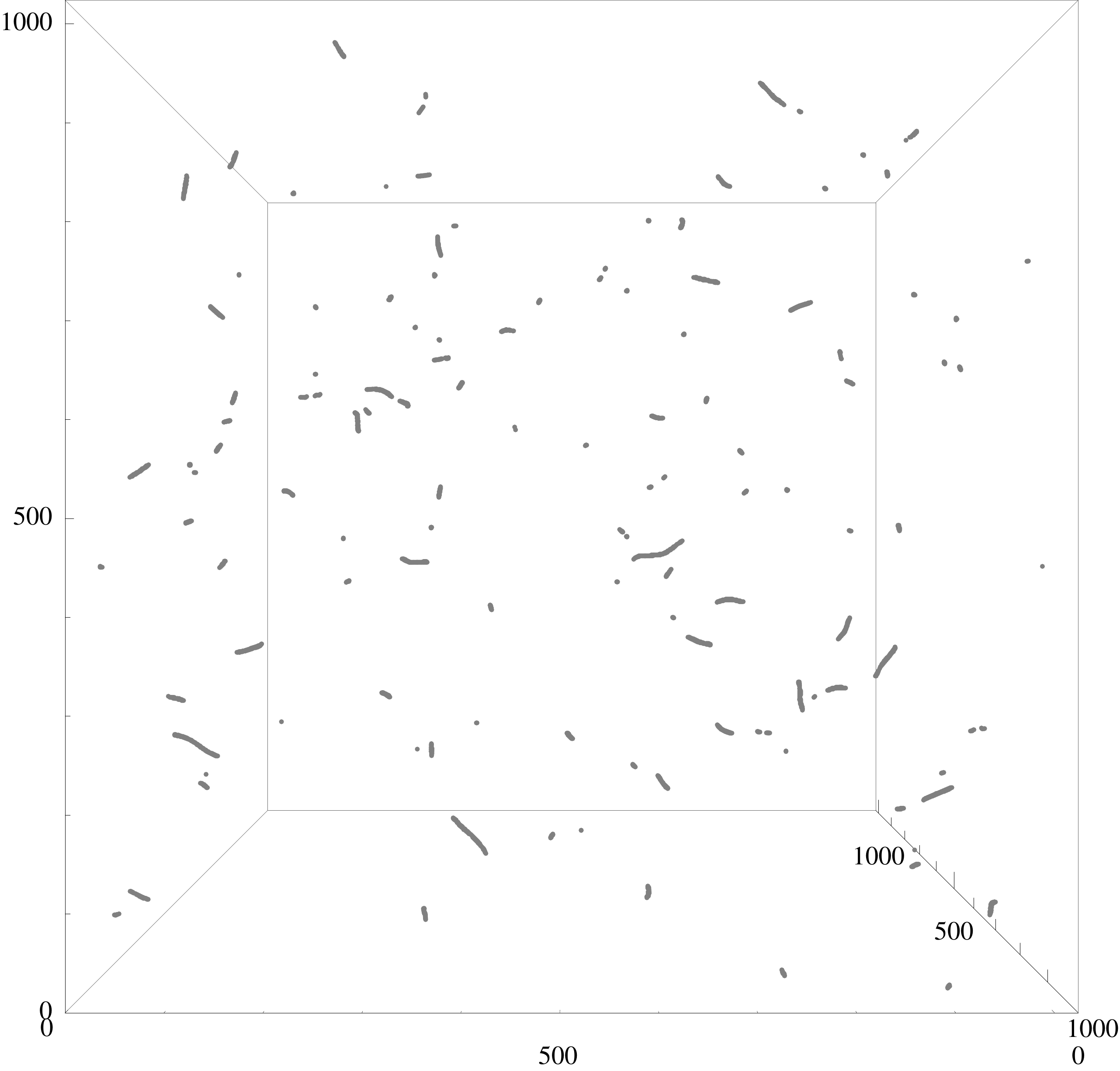}\\
\centering 
\includegraphics[width=0.49\textwidth]{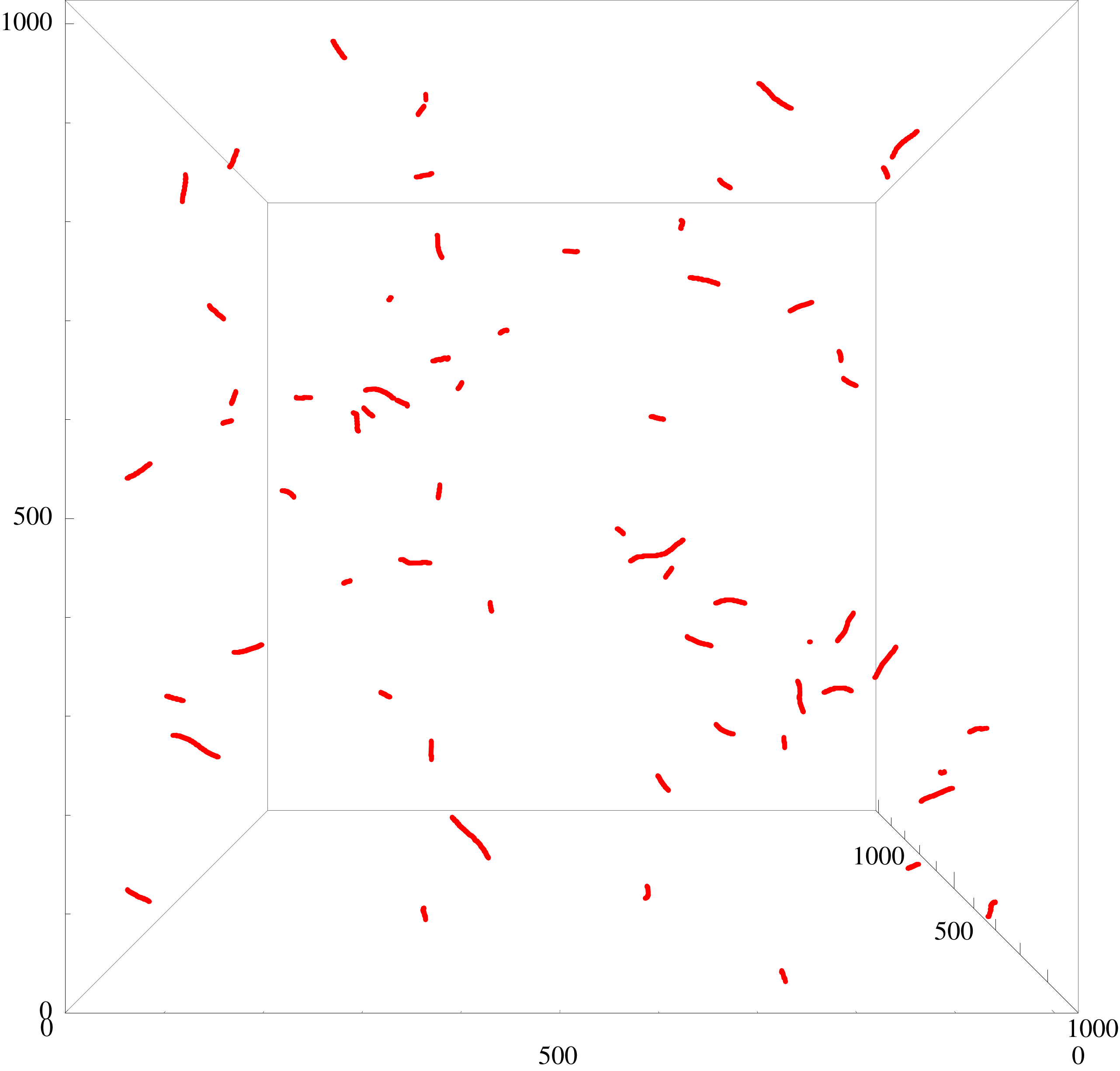}
\caption{Comparison of $pq$-segments calculated in 2 different manners for a $1024^3$ lattice. In the left panel of the upper part the winding representation is shown where segments are identified following the recipe given in Sec.~\ref{subsec_segmiden}. In the right panel, in turn, segments are found by the interaction potential. Grey regions correspond to points where the interaction potential is bigger than the threshold value $V_{\mathrm{int}}^{\mathrm{th}}=0.855$. In the lower panel every segment smaller than $20$ lattice-points has been discarded following the winding-based method.}
\label{fig:segments}
\end{figure}

In order to complement our segment localization method we perform an approximate consistency check. Besides winding identification, $pq$-strings can also be located using the interaction potential that is responsible for their formation. By inspection, we observed that $pq$-strings can be localized in the simulation box for values of the interaction potential greater than a threshold value of $V_{\mathrm{int}}^{\mathrm{th}}=0.855$.\footnote{It has to be noted that this is an approximated value motivated by trial and error inspections, enough for the purposes of confirming that our procedure works reasonably well, but  results might suffer small changes if this number is different.}  In the right panel of Fig.~\ref{fig:segments} we show an example of a simulation where we plot every point that exceeds that threshold value. This picture is accompanied by the corresponding winding identification method in the left panel of Fig.~\ref{fig:segments}. Comparing the output of both methods, we find a remarkable agreement between both approaches, which supports our segment identification treatment based on windings. We believe that the few differences between both pictures are caused by the uncertainty produced by the arbitrary choice of the threshold value.

Finally we have to discriminate between actual $pq$ double strings and small segments produced as a consequence of crossings of different $p$- and $q$-strings. Following the criterium set in \cite{Urrestilla:2007yw}, we get rid of crossings by removing segments smaller than $L_{\mathrm{AB}}=20$ in lattice units. Taking everything into account, the final result is shown  the left panel of Fig~\ref{fig:loops} and in the lower panel of Fig~\ref{fig:segments}. 


\section{Procedure to combine strings}
\label{sec_combine}

Interconnected string networks contain three different types of strings: individual $p$- and $q$-strings and composed bound states. In order to ensure bound string formation in field theories, an interaction term is typically included (last part of Eq.~(\ref{eq_potential})) which favors the individual string joining. Even though previous works have confirmed that bound strings can constantly form under the influence of this interaction term, it has been observed that their amount and length at the end of numerical simulations is low. Moreover, it appears that their lifetime is also relatively short indicating that they tend to unzip as the different $p$- and $q$- strings pull them in different directions.

One of the principal reasons for these discrepancies might be the limited dynamical range of numerical experiments, which is restricted by causality and computational resources. Initial conditions are typically set for the scalar and gauge fields of the system (or equivalently for their complex conjugates) and lead to the formation of the couple of ordinary Abelian gauge strings of the system. Therefore, in general, $pq$-strings are not present in numerical simulations from the beginning, instead one must wait until the formation of the bound states is energetically and dynamically favorable for the system. 

In order to enlighten and analyze these issues we explore a different technique to simulate interconnected string networks: {\it string combination}. The idea is to have a whole $pq$-string network interacting with an ordinary solitonic cosmic string network ($q$-strings in this case). In order to accomplish this, we create a new {\it fake} $pq$-string network superposing conveniently $p$- and $q$-strings. Essentially the $q$-string network is doubled and an extra set of $q$-strings placed on top of the already existing $p$-strings. Hence the whole $p$-string network is converted into the desired bound and heavier $pq$-string network. An illustrative example can be seen in Fig~\ref{fig:combine}, where we have included a snapshot of a string network before (left panel) and after (right panel) string combination. It can be seen that combination converts individual $p$-strings (in blue) of the left panel into $pq$-strings (in red) of the right one. String combination procedure introduces an innovative perspective into the analysis of the evolution of $pq$-bound states. Using an artificial initial configuration we will be able to determine if the system is comfortable with a high amount of heavy strings or on the contrary it prefers to break them into small pieces and decrease their relative relevance.

The procedure must be applied when the Abelian Higgs strings of the system are already formed. It would be meaningless to try to superpose fields before strings are formed, since it would not lead to a string combined scenario. This technique to combine different strings is somewhat different to that used in \cite{Rajantie:2007hp}. In that work the authors used a totally aligned phase distribution for the scalar fields to begin the simulation with a totally formed $pq$-string network.  Given the symmetry of the equations of motion of our model, such initial conditions  would lead us to a totally indistinguishable pair of AH cosmic strings evolving in the same way, that would not reproduce the evolution of interconnected networks.  That is why, in our case,  we apply the combination procedure to strings that are already formed, and  the initial $pq$-network of our combined simulations coexist with another {\it single} string network.
\begin{figure}[h!]
\centering
\includegraphics[width=0.85\textwidth]{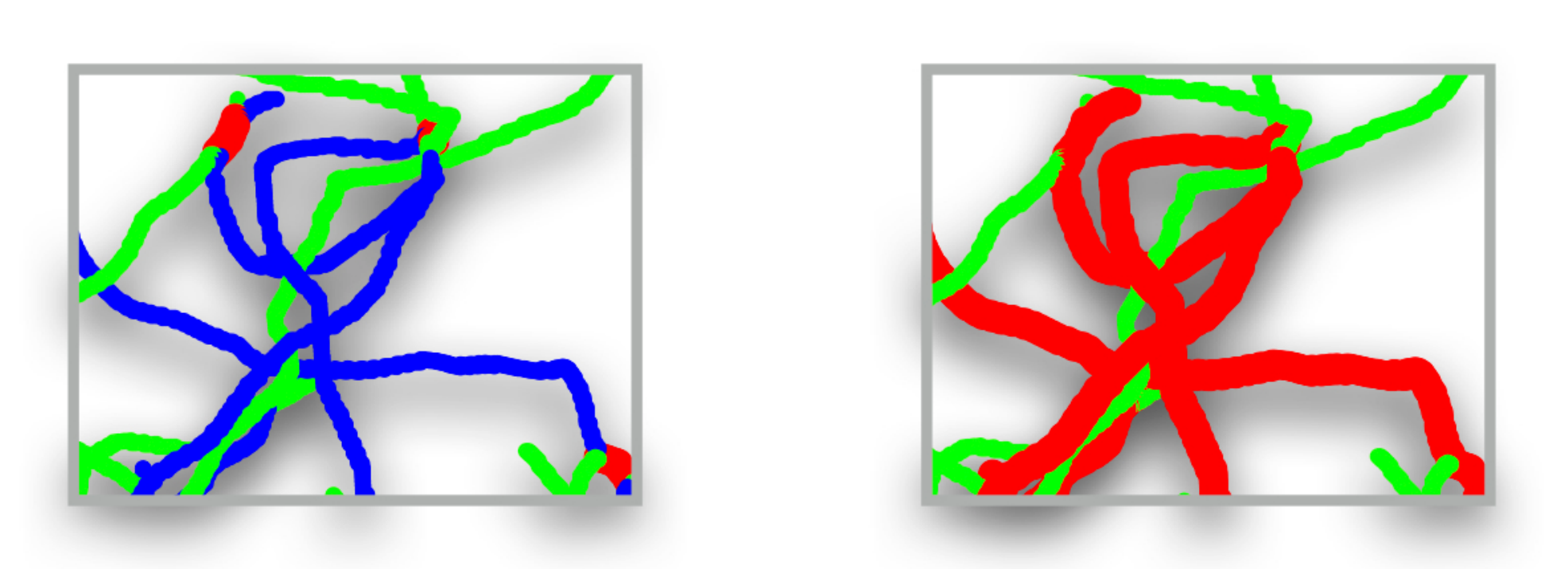}
\caption{Comparison of a patch of the simulation box before (left) and after (right) string combination. I n the left panel, it can be seen that some red points are accidental crossings, and not $pq$-segments proper. It is clear that  $p$-strings (in blue) of the left panel are completely converted into $pq$-strings (red strings) after combination.}
\label{fig:combine}
\end{figure}

Step by step, the combination mechanism applied to fields can be summarized as follows: the system before combination of strings is composed by two (nearly) independent complex scalar fields, $\phi$ and $\psi$, set at their corresponding vacuum and with their corresponding strings ($p$ and $q$ respectively). String combination procedure leaves the first complex scalar field $\phi$ and its canonical conjugate unaltered:

\begin{eqnarray}
\phi \rightarrow \phi = |\phi|e^{i\theta_{\mathrm{A}}}\, , \\
\Pi_{\mathrm{A}} \rightarrow \Pi_{\mathrm{A}}\, ,
\end{eqnarray}
the gauge field associated to $\phi$ also remains unaltered.

The second field $\psi$, however, is substituted by a precise combination of both initial scalar fields:

\begin{equation}
\psi \rightarrow \psi = |\phi||\psi| e^{i(\theta_{\mathrm{A}}+\theta_{\mathrm{B}})}\, .
\end{equation}

This new scalar field vanishes when $\phi$ and/or $\psi$ are zero, \ie the strings associated to this new combined scalar field are located wherever original $p$- and $q$-strings are present. In some sense, we double the $q$-string network by adding a new set of $q$-strings at the position of the original $p$-strings.

The canonical conjugate of $\psi$ must also be modified:

\begin{equation}
\Pi_{\mathrm{B}} \rightarrow \Pi_{\mathrm{B}} = \phi \Pi_{\mathrm{B}} + \psi \Pi_{\mathrm{A}}\, .
\end{equation}

Finally, the new gauge field is the linear superposition of the original gauge fields (and similarly with the time derivative of the gauge field $E_{\mathrm{B} i}$):

\begin{eqnarray}
\textbf{B}_{i} \rightarrow \textbf{B}_{i} = \textbf{B}_{i} + \textbf{A}_i\, ,\\
\textbf{E}_{\mathrm{B} i} \rightarrow \textbf{E}_{\mathrm{B}i} = \textbf{E}_{\mathrm{B}i} + \textbf{E}_{\mathrm{A}i}\, .
\end{eqnarray}

We have observed that the evolution and the scaling dynamical range are optimized when the string combination is performed at the end of the dissipative phase, at $\tau_{\rm comb}=105$, but before the beginning of the core growth phase. On the one hand, we let the system remove the energy excess caused by the random initial conditions; on the other, we ensure that at the combination strings are already formed.


\section{Results}
\label{sec_results}

\subsection{Network properties}
\label{network}

In this section we present the most relevant results obtained in our simulations. We performed 5 matter plus 5 radiation realizations for both combined and normal simulations, using the parameters values as described in Sec.~\ref{sub:simdet} and for $\kappa=0.9$ (later we also report on simulations with $\kappa=0.95$). The analysis will be mainly focused on combined simulations in the matter era (the results in radiation domination are similar, and will be shown later), where one of the complex scalar fields of the model has been substituted by a combination of both fields in the initial stages of the simulation, in order to produce a whole network of bound $pq$-strings. We will explore the effects produced by these new initial conditions in the amount and evolution of different types of string and the new results will be compared with those obtained from {\it normal} simulations, where string combination has not been utilized. In addition, average velocity distributions of different networks will be analyzed. Velocities of interconnected networks, and specially velocities of $pq$-strings, have been measured for the first time using field theory based velocity estimators. We will show as well as compare results of different type of simulations.

Any reasonable and useful description of string networks must be derived from simulations evolving in the scaling regime. Scaling ensures cosmological viability of such objects and enables the extrapolation of the results to scales of cosmological interest. In the scaling regime, the characteristic lengths of the network evolve proportionally to the cosmic time, \ie they grow linearly with the horizon. In the specific case of interconnected $pq$-networks, the scale invariant evolution was confirmed by different works \cite{Urrestilla:2007yw,Sakellariadou:2008ay}. In order to confirm scaling of our different simulations, we calculate the typical length of the strings using  the following two different methods:

The string separation or the characteristic length of the network is typically defined in terms of a reference volume $V$ and the length of the strings within it $L$,

\beq
\xi = \sqrt{\frac{V}{L}}\, .
\eeq

One way to derive $L$ is to measure the length of each string adding up the number of plaquettes pierced by them. This is the typical method and what we use to study the scaling of the different ingredients of our model and which we call $\xiW$. However, when the object to be explored is the whole system, one can estimate that value using field theoretical estimators such as the Lagrangian, with  

\beq
L = -\mathcal{L}V/\mu\, ,
\eeq
where $\mu$ is the string tension. The string separation determined by this method will be represented as $\xiL$.

Figure~\ref{fig_xiLW} shows characteristic lengths of the whole system measured using the two different methods mentioned in previous lines: Lagrangian method $\xiL$ (in black) and winding method $\xiW$ (in purple). The left panel shows the results for combined simulations, while the right panel shows the results for the normal case. The normal simulation achieves the scaling regime  for $\tau\geq150$. Remarkably, a similar result is obtained for the combined case, which reaches the scale invariant evolution more or less at the same simulation stage. A sudden step in $\xiW$ around $\tau\sim100$, just before the start of the core growth phase, represents the string combination moment. It does not seem to affect the latter evolution of the network towards scaling.

\begin{figure}[h!]
\centering
\includegraphics[width=0.49\textwidth]{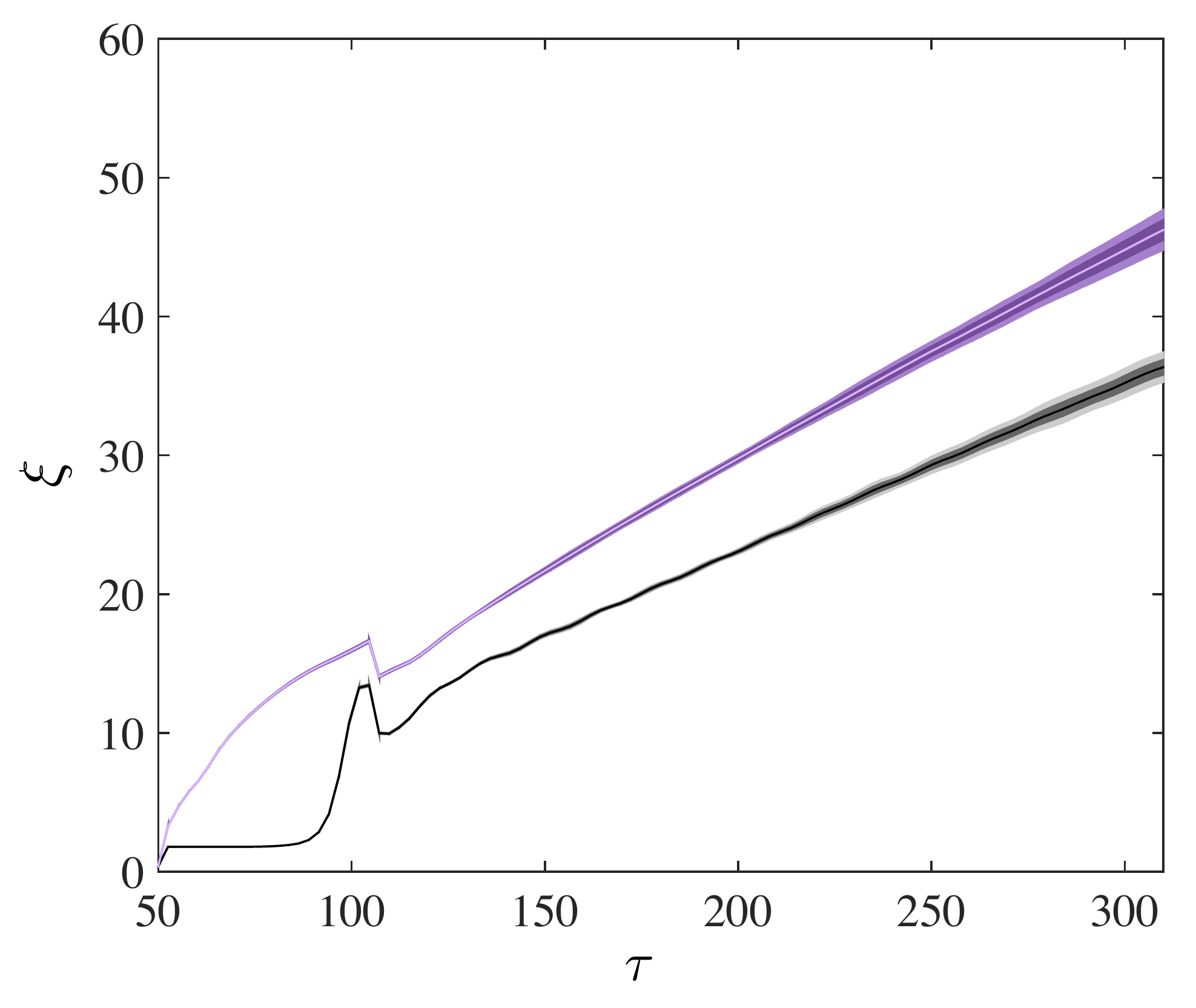}
\includegraphics[width=0.49\textwidth]{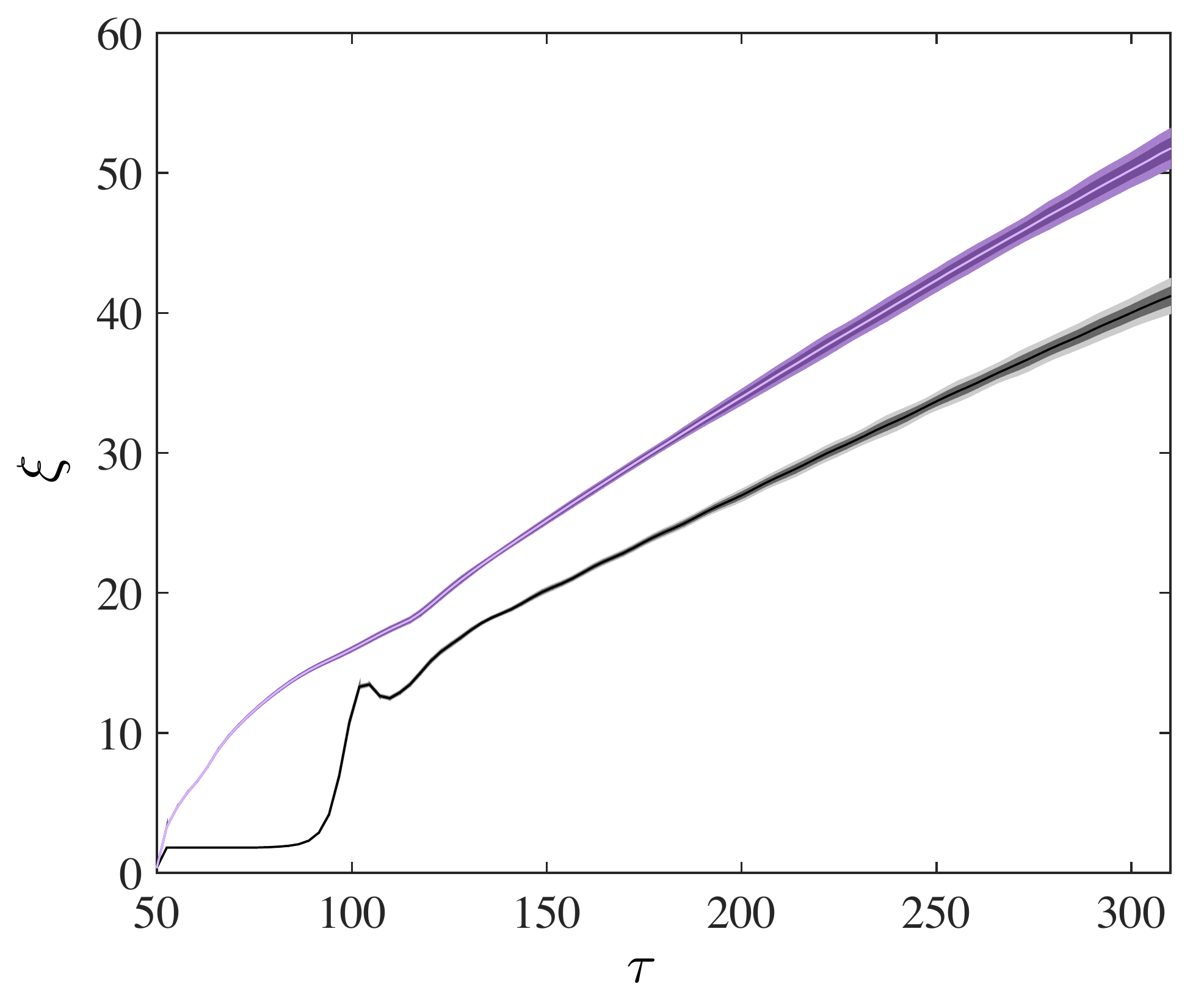}
\caption{Characteristic string length, $\xi$, measured using the Lagrangian (black line) and windings (purple line) for combined (left) and normal (right) simulations in the matter era. Shaded regions correspond to $1\sigma$ and $2\sigma$ confidence limits obtained averaging over 5 realizations.}
\label{fig_xiLW}
\end{figure}

Interconnected string systems, however, are composed by three different type of strings. Hence a full confirmation of the scaling of the network requires further analysis and the study of the behavior of each individual set of strings. We start analyzing the behavior of individual $p$- and $q$-string systems, \ie networks composed of loops or segments that are not forming $pq$-bound segments. We define the string separation for such cases as,

\beq
\xiW^{\rm p} = \sqrt{\frac{V}{L_{\rm p}}}\, , \quad \quad \xiW^{\rm q} = \sqrt{\frac{V}{L_{\rm q}}}\, ,
\eeq
where $L_{\rm p}$ and $L_{\rm q}$ are the total length of $p$- and $q$-strings where the length corresponding to $pq$-segments has been subtracted.

Figure~\ref{fig_xiLS} includes the curves associated to those quantities: the blue colored region represents $\xiW^{\rm p}$ ($p$-strings), whereas the green is for $\xiW^{\rm q}$ ($q$-strings). It has to be noted that the starting point of the axis in this case is different, while global statistics of the system (total $L$, $\mathcal{L}$...) are collected for the whole simulated time, $pq$-segment identification procedure, and hence the calculation of $L_{\rm p}$ and $L_{\rm q}$, is only applied in the core growth phase, \ie $\tau\geq114$.

The blue line of the combined simulation reflects clearly that the $p$-string network is far from being scaling in the initial stages. The system approaches gradually the linear regime and it is achieved approximately at $\tau\sim200$. On the contrary, individual string networks in the normal case appear to be scaling in the whole measured range. The difference between the blue and green curves in the left panel comes from the fact that the field $\psi$ has been doubled and an extra set of $q$-strings created on top of the already existing $p$-strings. Hence, the initial length of $p$-strings is approximately zero $L_{\rm p} \sim 0$ (there is no free $p$-strings) and thus $\xiW^{\rm p}$ enormous. Nevertheless, the profile of the blue curve suggests that as time goes by the total length of $p$-strings increases considerably until the system starts following the linear dependence at $\tau\sim200$.

\begin{figure}[h!]
\centering
\includegraphics[width=0.49\textwidth]{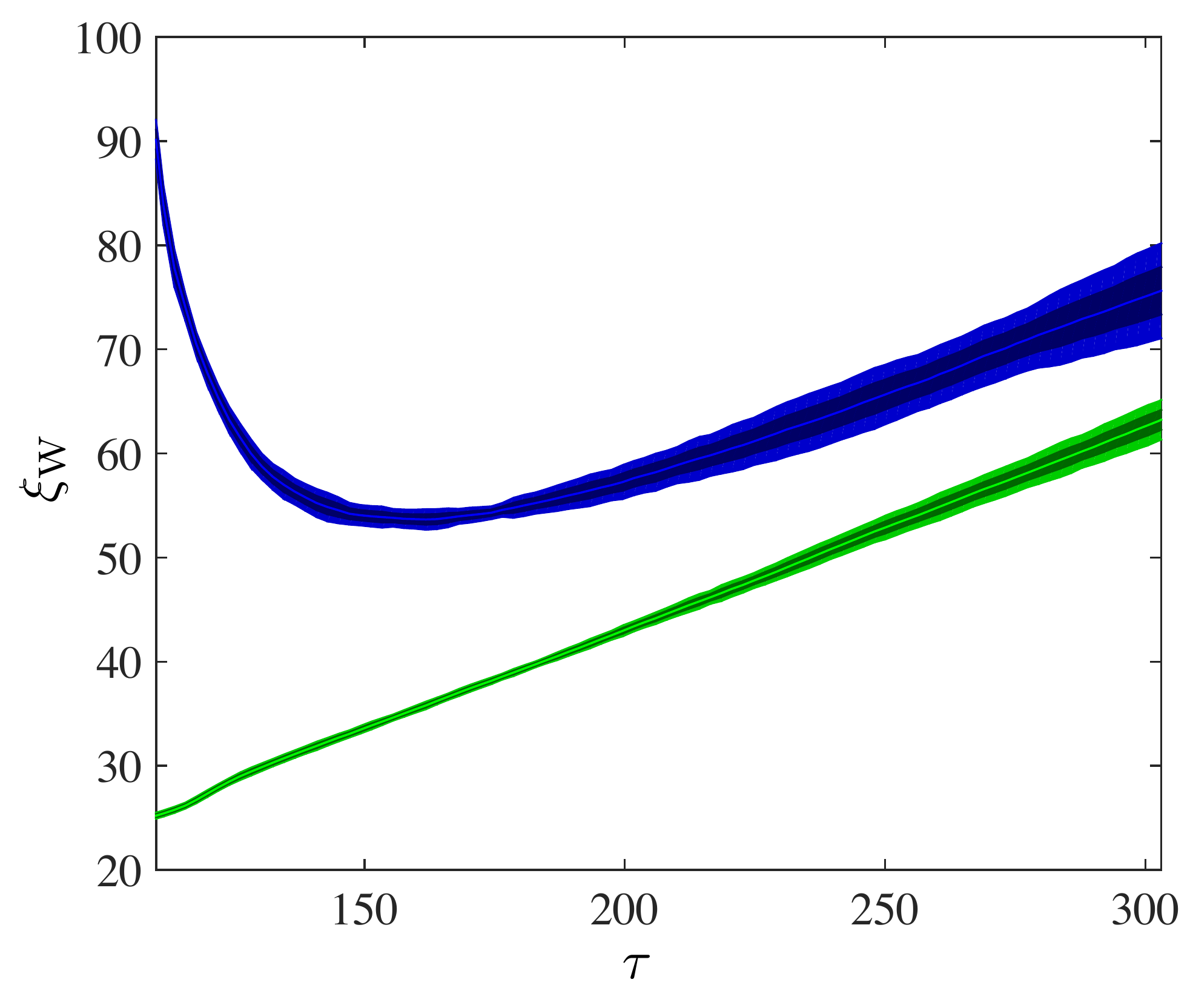}
\includegraphics[width=0.49\textwidth]{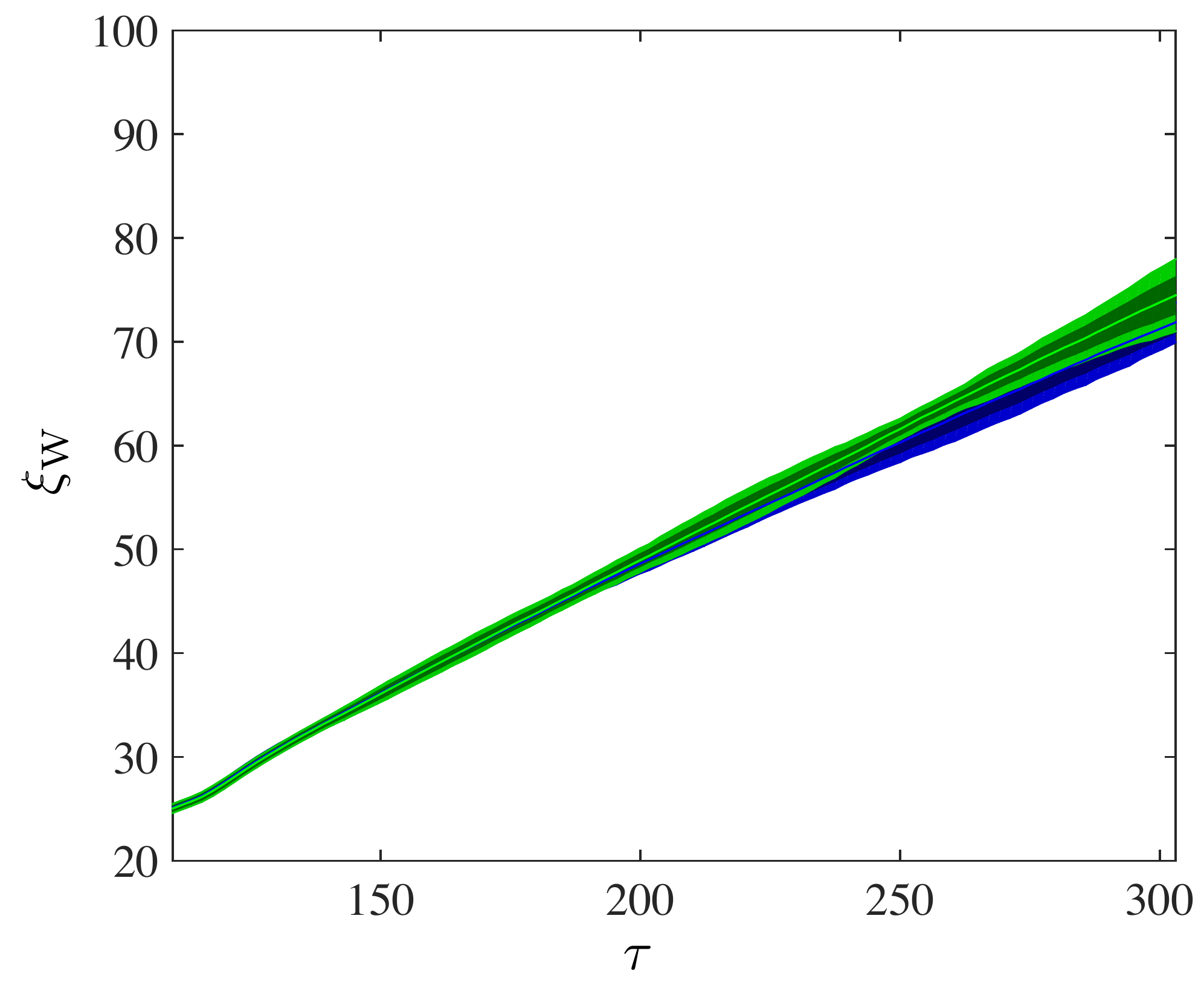}
\caption{Characteristic length $\xiW$ of individual $p$-  (blue) and $q$-strings (green) measured using string windings for combined (left) and {\it normal} simulations (right). Shaded regions correspond to $1\sigma$ and $2\sigma$ confidence limits obtained averaging over 5 realizations.}
\label{fig_xiLS}
\end{figure}

It is evident that the $p$ network needs a relaxation period after string combination, where part of the total length in bound states is transferred to $p$-strings. Such a period is clearer observed and understood analyzing the evolution of the $pq$-string fractions. We define two different fractions:

\begin{enumerate}
	\item We measure the ratio of the total length of $pq$-strings to the length of the individual $p$-strings:
	\beq
	\fp = \frac{L_{\rm pq}}{L_{\rm p}}\, ,
	 \eeq
	where $L_{\rm pq}$ is the sum of the length of all $pq$-segments.
	
	\item Alternatively we want to measure what the fraction of the total length of $pq$-segments is with respect to the total length of the whole system:
	
	\beq
	\ftot = \frac{L_{\rm pq}}{L_{\rm p} + L_{\rm q} + L_{\rm pq}}\, .
	\eeq
\end{enumerate}

These two magnitudes are represented in Figure~\ref{fig_fAB}, $\fp$ in blue and $\ftot$ in black. The figure corresponding to the normal simulation shows what previous works found: the total length of the $pq$-strings as compared to the total length of the whole system is really small, giving $\ftot\sim0.02$ \cite{Urrestilla:2007yw}. The blue line in this case is approximately twice the grey one because the lengths of $p$- and $q$-strings contribute at the same level to the total. 

The trend exhibited by the string combined case is very different. Initially $\fp$ is forced to be 1 by the combination process, that is, all $p$-strings are in $pq$ states. However, the curve followed by this fraction shows that most of the length of $pq$-strings is converted into $p$-string's length. In other words, the system seems to not feel comfortable with such a high amount of bound states and wants to break them. For instance, by the time the network has finished the relaxation period $\tau\sim200$, the fraction decreases to $\fp\sim0.4$. The evolution of the individual strings seems to provoke the unzipping of the bound states. The final evolution of the fraction after the relaxation period is smoother and approaches an asymptotic value of $\ftot\sim0.05$, which is roughly twice the quantity observed in normal simulations.

\begin{figure}[h!]
\centering
\includegraphics[width=0.49\textwidth]{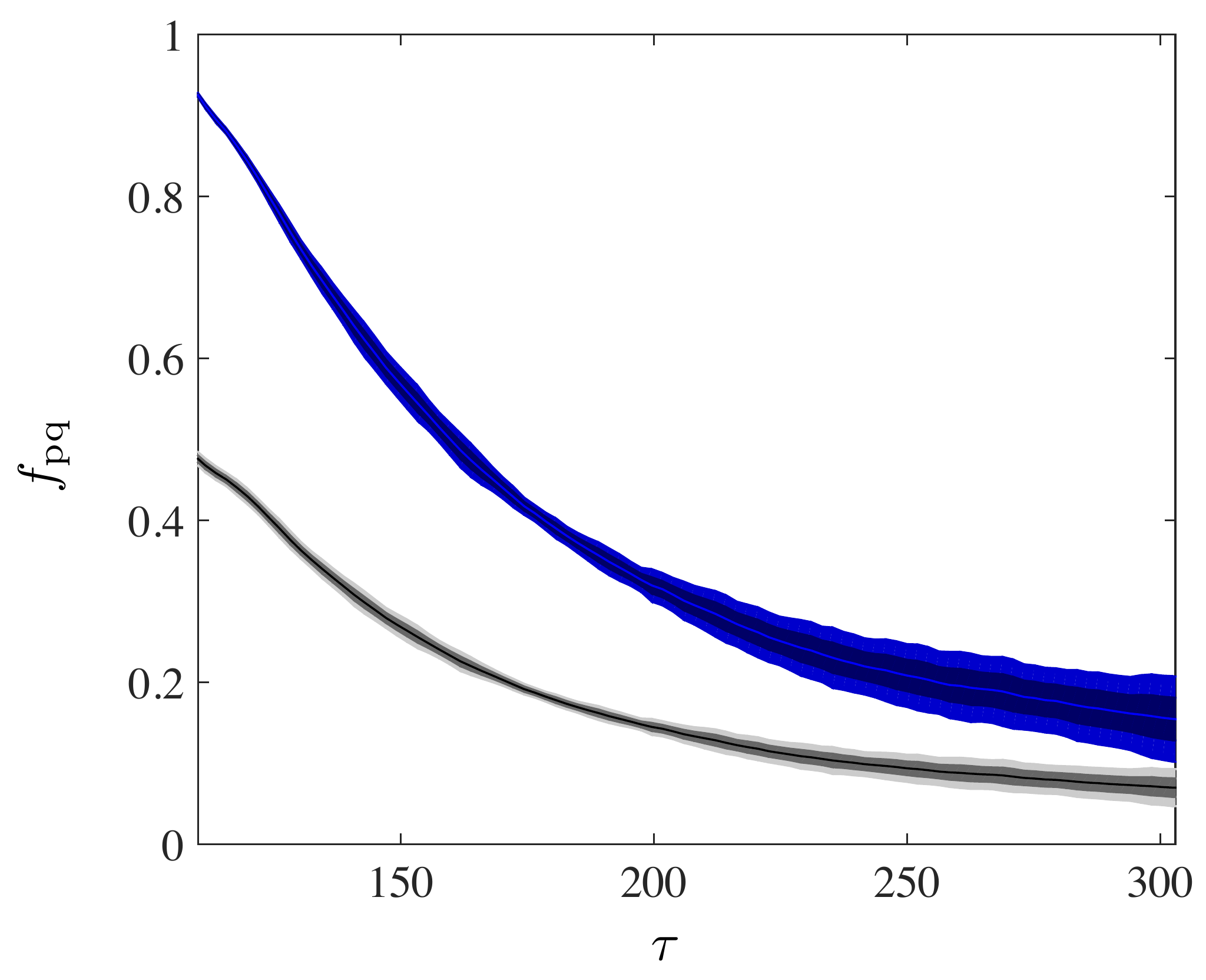}
\includegraphics[width=0.49\textwidth]{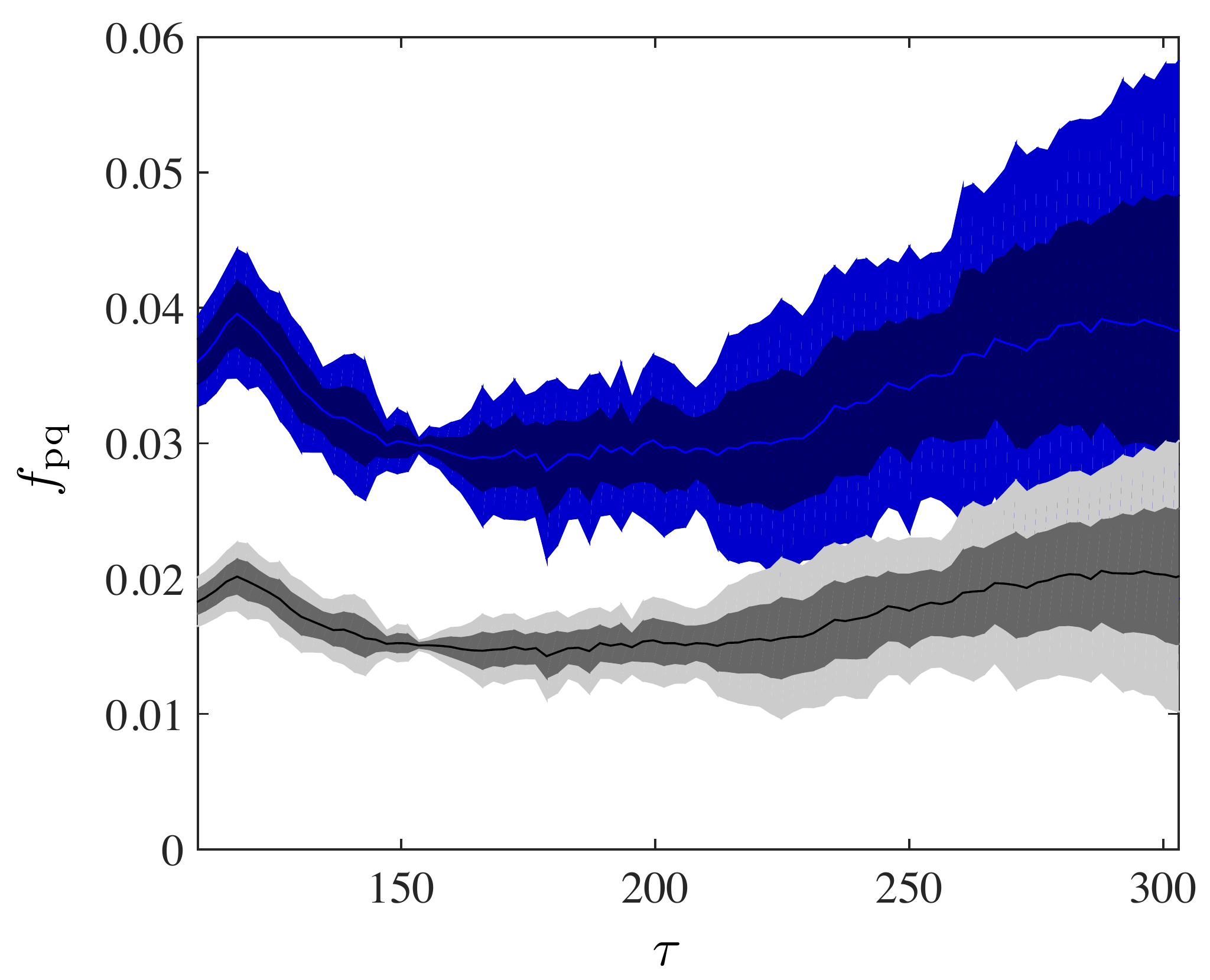}
\caption{$pq$-string fraction in combined (left) and normal simulations (right). The blue line corresponds to $\fp$, while the black corresponds to $\ftot$. Shaded regions represent 1$\sigma$ and 2$\sigma$ errors obtained by averaging over 5 different realizations.}
\label{fig_fAB}
\end{figure}

Scaling of the $pq$-segments has also been explored separately. We define the correlation length of $pq$-strings in the following way:

\beq
\xiW^{\rm pq} = \sqrt{\frac{V}{L_{\rm pq}}}\, .
\eeq

Figure~\ref{fig_pqscaling} shows that scaling is remarkably well achieved in simulations where strings have been combined. Furthermore, the linear proportionality is significantly better in this case than in normal simulations, where the characteristic length could barely be approximated to a straight line. The linear regime is reached at around $\tau\sim200$, which is consistent with the relaxation behavior exhibited by the system in other observables. If we compare both simulations, one can see that the correlation length of bound states in the combined case is smaller than in normal simulations, which indicates that the $pq$-strings are in general more and longer in combined simulations. In both cases these correlation lengths are much bigger than the functions shown in Fig.~\ref{fig_xiLS}. 

\begin{figure}[h!]
\centering
\includegraphics[width=0.49\textwidth]{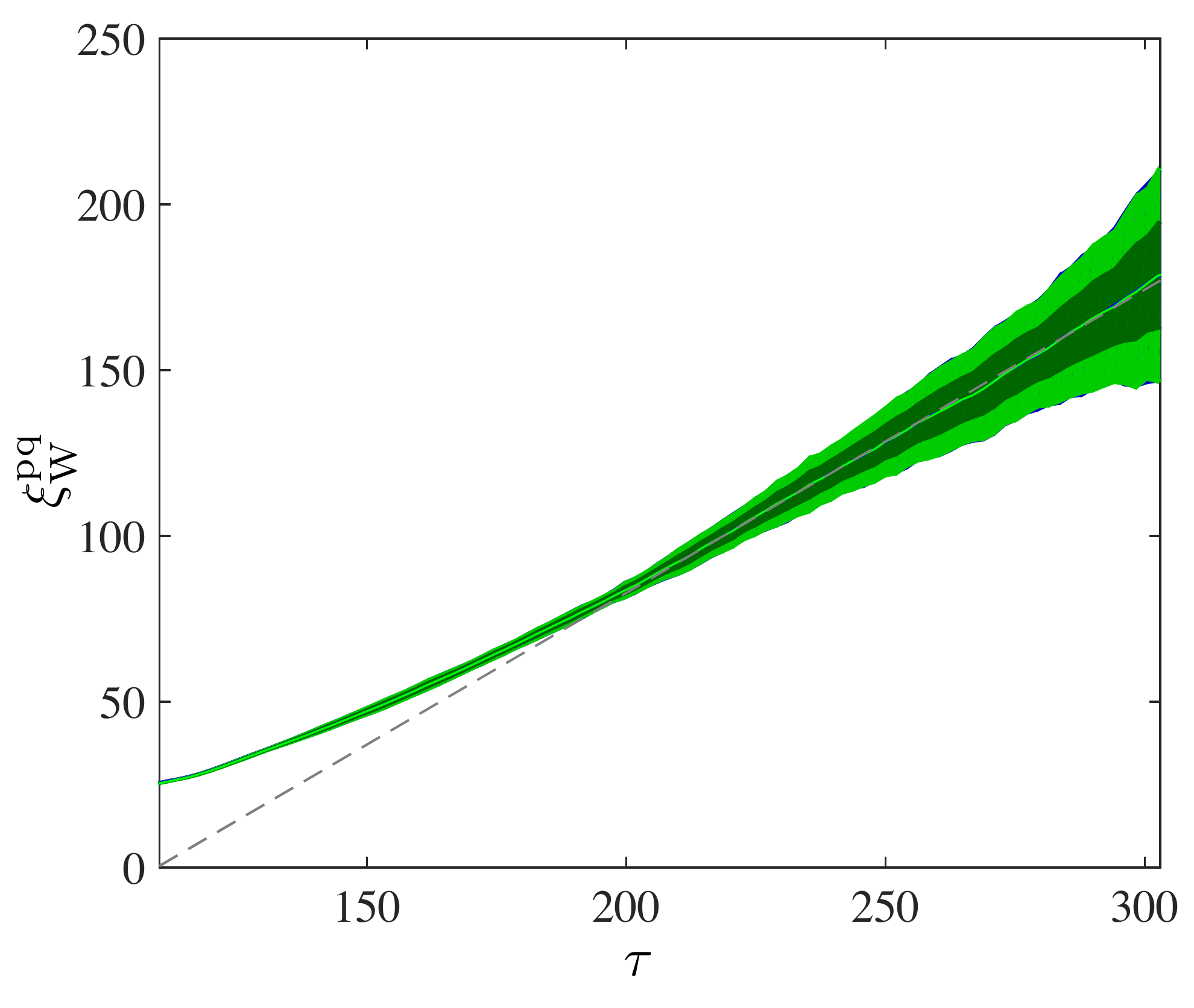}
\includegraphics[width=0.49\textwidth]{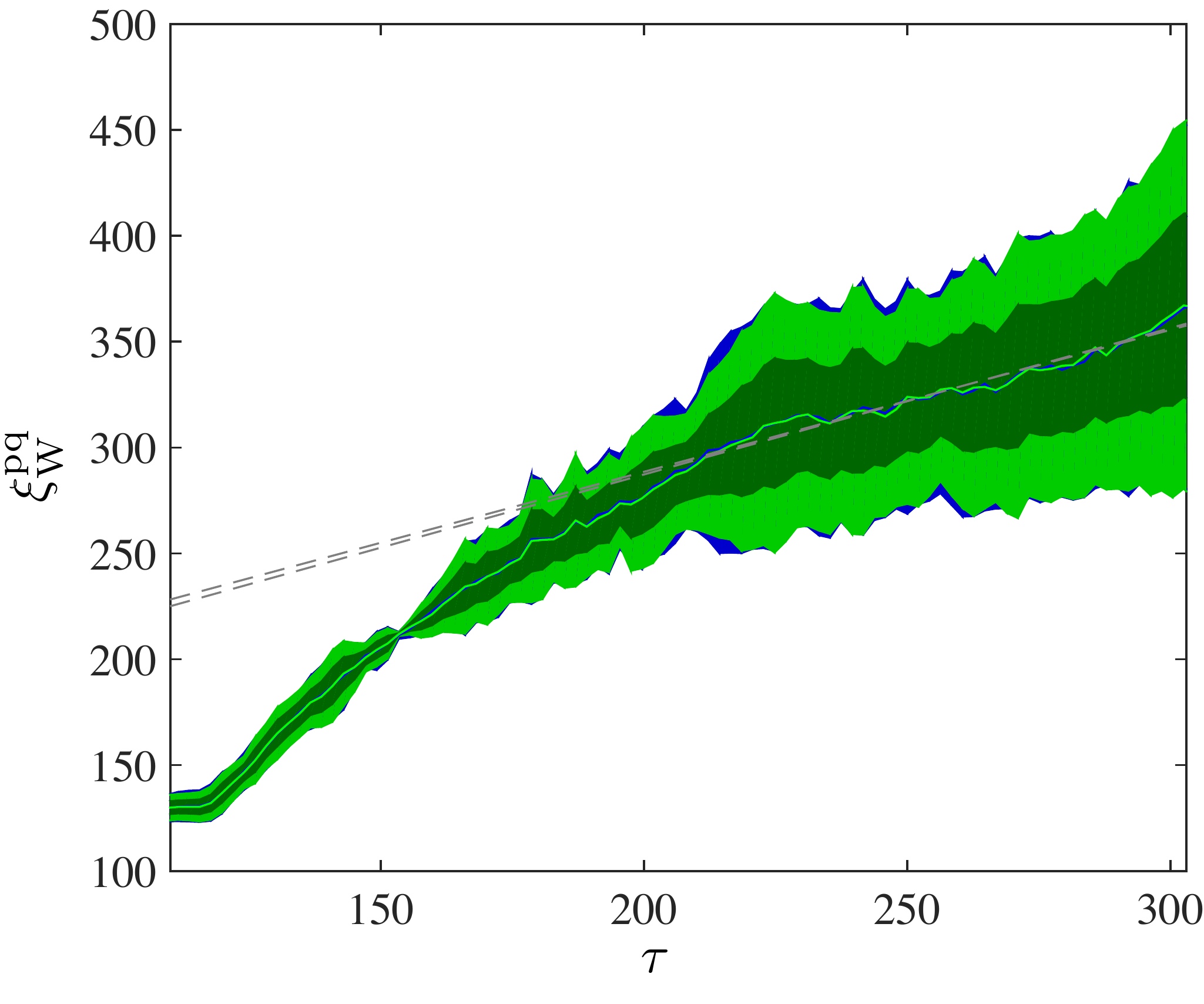}
\caption{Characteristic length of $pq$-segments $\xiW^{\rm pq}$ measured using windings for string combined (left) and normal simulations (right).}
\label{fig_pqscaling}
\end{figure}

\begin{figure}[h!]
\centering
\includegraphics[width=0.49\textwidth]{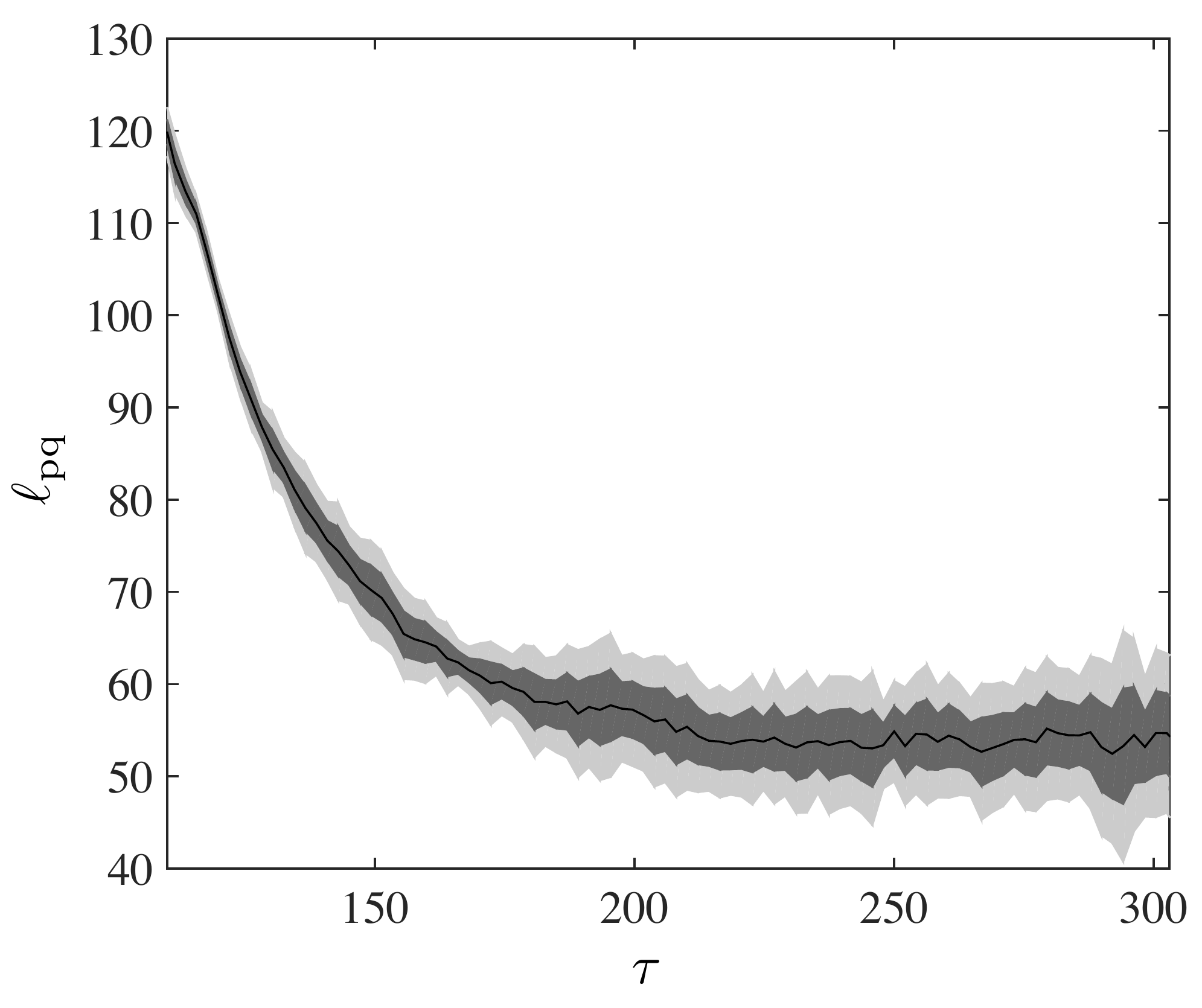}
\includegraphics[width=0.49\textwidth]{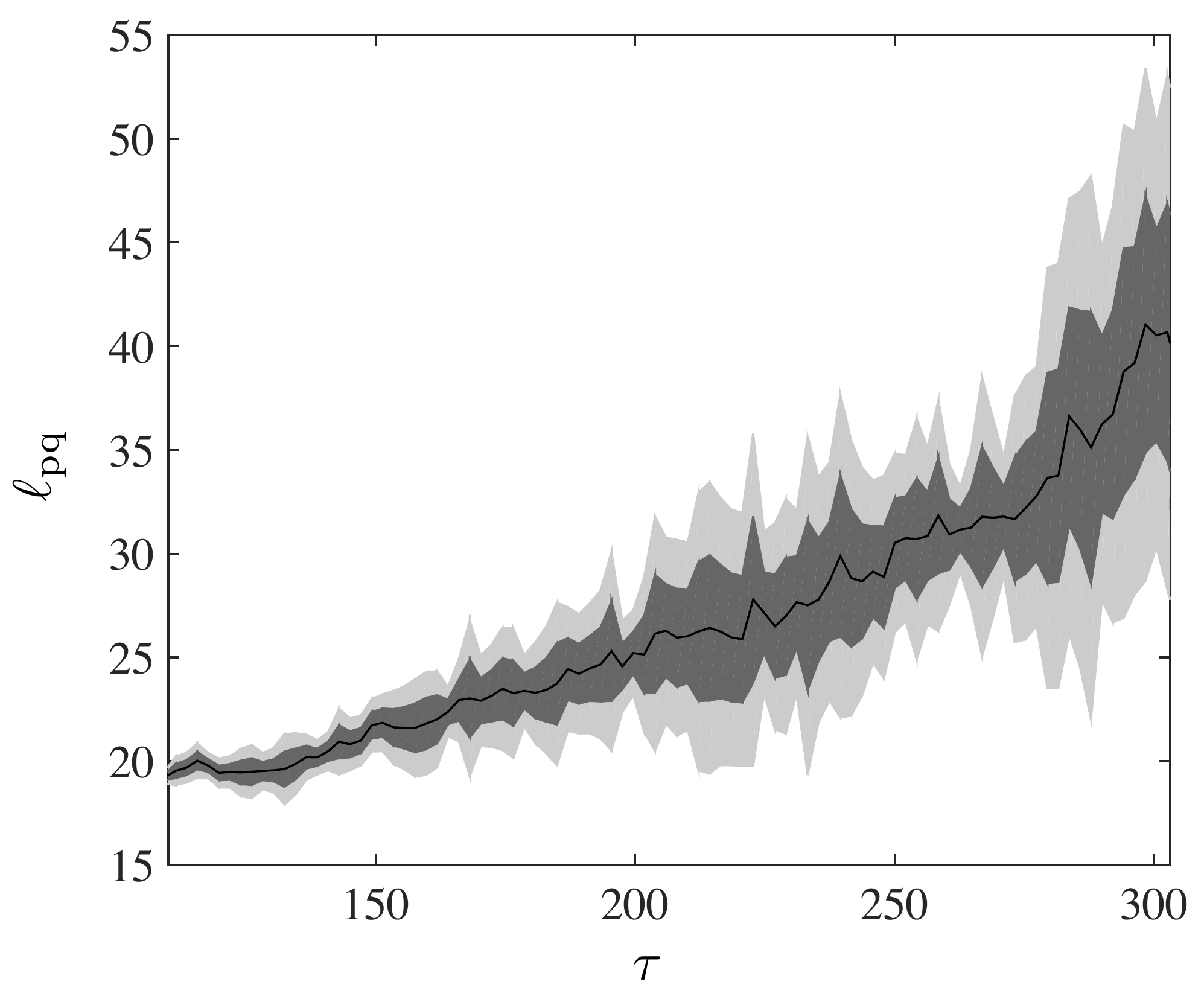}
\caption{$pq$-string average physical length $\lpq$ in combined (left) and normal simulations (right). Shaded regions correspond to 1$\sigma$ and 2$\sigma$ errors obtained by averaging over 5 different realizations.}
\label{fig_pqAvLength}
\end{figure}

Another interesting characteristic length of the $pq$-bound segments its their average physical length. It is defined in the following manner:

\beq 
\lpq = \frac{L_{\rm pq}\Delta x}{N_{\rm pq}}\, ,
\eeq
where $N_{\rm pq}$ is the total number of $pq$-segments. Figure~\ref{fig_pqAvLength} shows $\lpq$ for both type of simulations. The profile of the curves is very different: whilst the average length in the normal case is approximately a linearly increasing function of time, the average physical length of the $pq$-strings in combined simulations tends to an asymptotic constant value. The asymptote is located at $\lpq\sim55$ in the matter case and the curve is almost flat in the period of time after the system relaxation, \ie $\tau\geq200$. Interestingly, the average length in normal simulations remains below that value. The scale invariant evolution is generally better and acquired faster by combined simulations, where not only the whole system scales, but also $p$, $q$ and $pq$-networks separately do. One could expect that bigger simulations of the normal case with a larger dynamical range would tend towards the asymptotical evolution depicted by combined simulations. 

We have also performed simulations in radiation dominated scenarios. We observe that in general the results extracted from matter dominated and radiation dominated cosmologies are very similar and no significant distinction can be made between both cases. 

\begin{figure}[h!]
\centering
\includegraphics[width=0.60\textwidth]{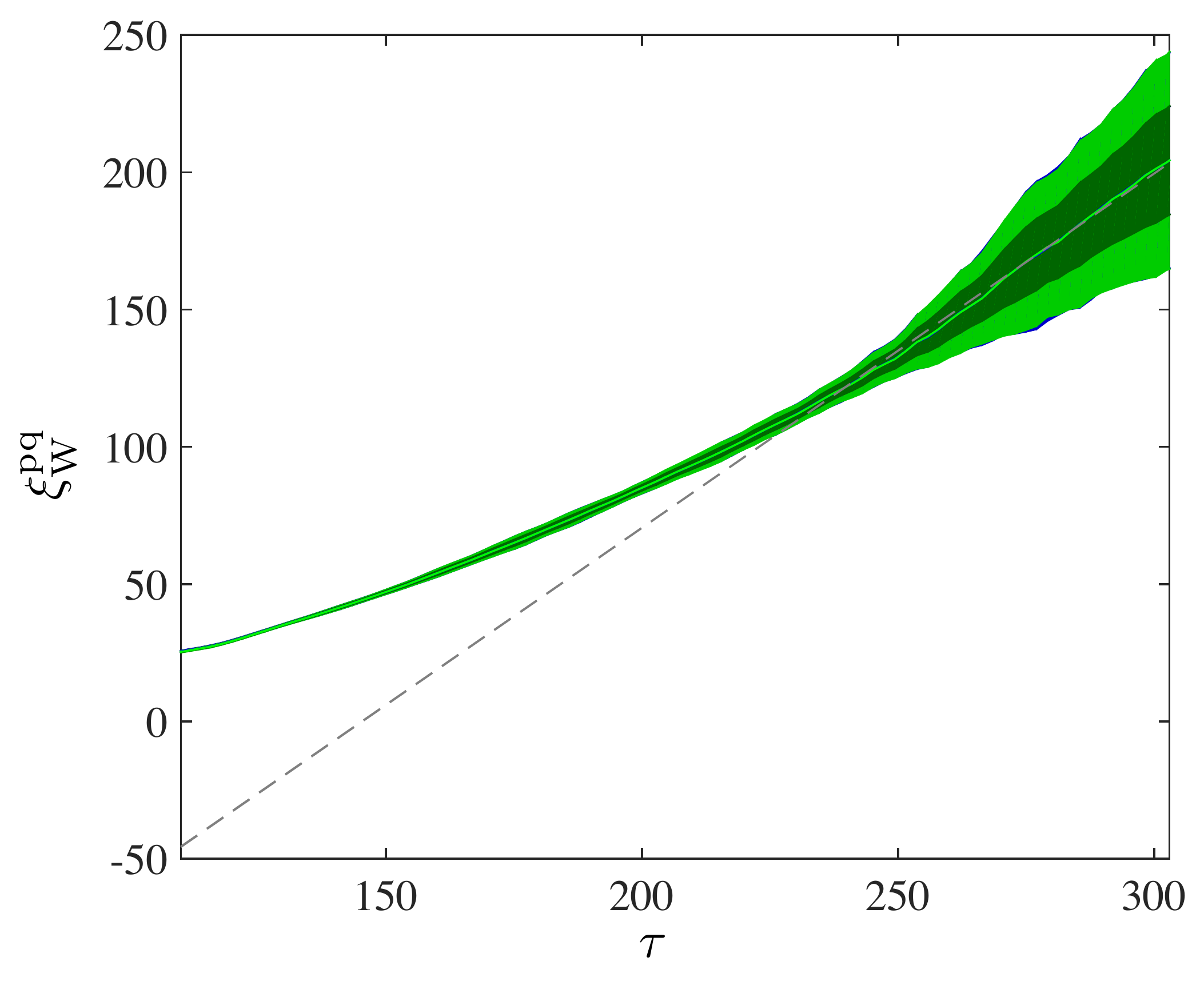}
\caption{Correlation length of $pq$-strings in combined simulations in the radiation domination era.}
\label{fig_pq_xi_rad}
\end{figure}

\begin{figure}[h!]
\centering
\includegraphics[width=0.60\textwidth]{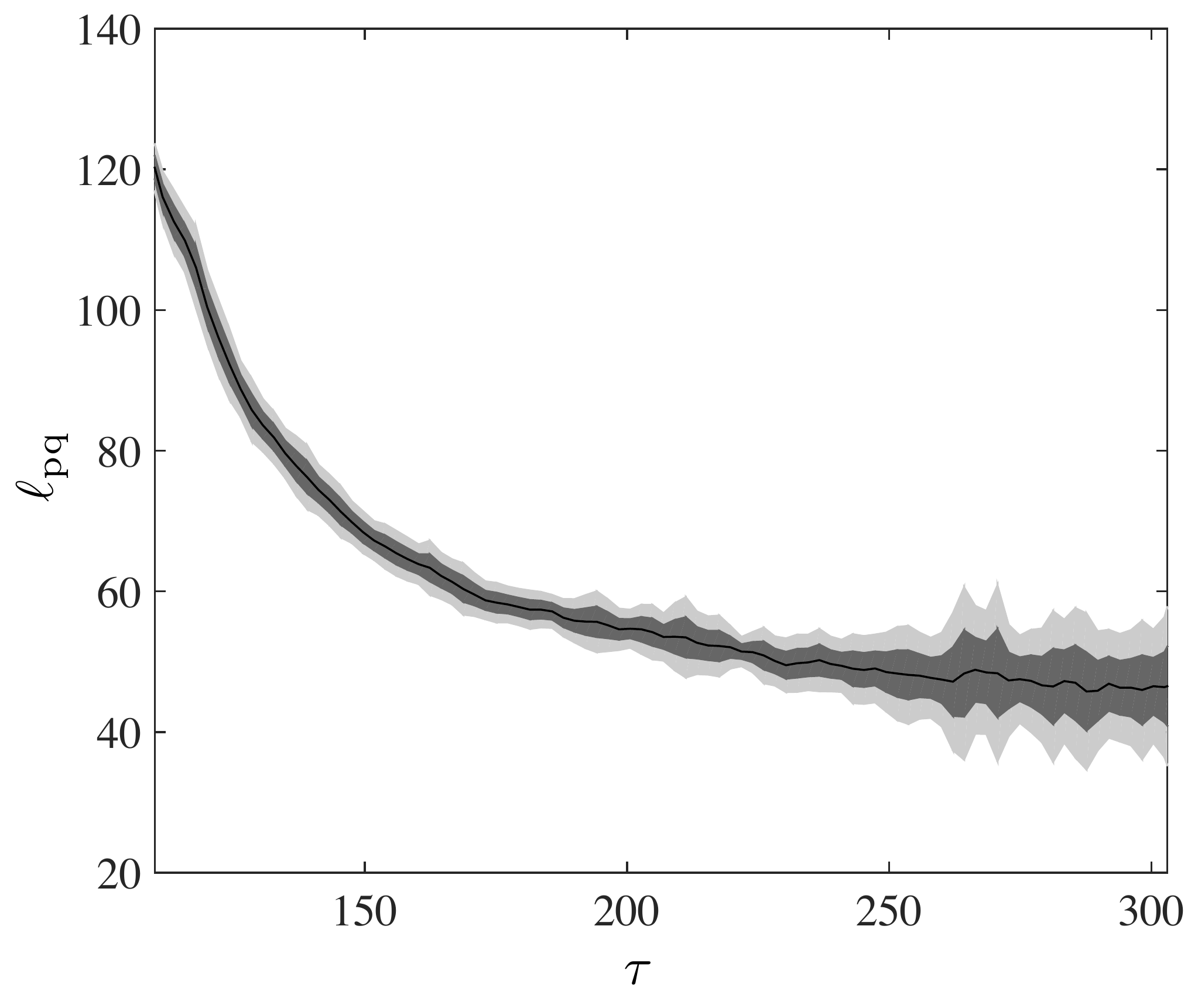}
\caption{Average physical length of $pq$-strings in combined simulations in radiation domination era.}
\label{fig_pq_rad}
\end{figure}

The differences come mostly from the lower damping in the equations of motions produced by the different expansion rate of universes evolving in radiation domination. The lower damping term affects the relaxation period of string combined simulations. Figs.~\ref{fig_pq_xi_rad} and \ref{fig_pq_rad} show $\xiW^{\rm pq}$ and the average physical length of $pq$-strings in radiation domination eras. As it can be seen, both pictures reproduce essentially the evolution depicted in the matter case, but with the difference that the scale invariant evolution is acquired later. Scaling of the bound segments is also observed in radiation domination era and the scaling regime is achieved at $\tau\sim 225$. As it happened in matter domination, the scaling exhibited by combined simulations is better than the behavior of the normal simulations.

The average physical length also tends towards an asymptotic value, as Fig.~\ref{fig_pq_rad} shows. However, as with the scaling, the network in radiation domination needs more time to relax and the asymptote is reached later $\tau\sim250$. The value for this case $\lpq\sim45$  is slightly below the asymptotic value obtained in matter domination.

Finally, we have also modified the coupling parameter of the interaction potential responsible for the $pq$-string formation, $\kappa$, so as to explore its relevance in the creation, evolution and amount of bound states. We increased its value to $\kappa=0.95$ and performed simulations in both normal and combined scenarios. In fact, such value of $\kappa$ is really close to the maximum value allowed for this model, as it is in indicated in Eq.~(\ref{eq_klimit}). 

On the one hand, our normal simulations confirm previous results of \cite{Urrestilla:2007yw}: increasing the value of the coupling constant from $\kappa=0.9$ to $\kappa=0.95$ only produce a marginal increase of the fractions of the bound states and in general did not produce remarkable changes in the evolution of them. On the other hand, combined simulations with $\kappa=0.95$ show that, despite the higher coupling constant of the attractive potential term, the overall dynamics of the system tends to split bound states and as in the case with $\kappa=0.9$ their relative fraction decreases rapidly. In terms of numbers, we only observe minimal changes, which translate into slightly bigger values of the asymptotic length of the bound states, $\lpq\sim70$, and asymptotic total fraction $\ftot\sim0.12$, in both matter and radiation eras.

\subsection{Velocity estimation}
\label{subsec_vel}

Average velocities of the network have been measured using the gauge invariant field theoretical velocity estimators proposed in \cite{Hindmarsh:2008dw,StringProPaper}, which exploit the fact that the electric field and canonical momentum of the scalar field in moving strings can be obtained by boosting the static field distribution. The velocity estimators have the following form:

\beq
 \vf = \frac{\textbf{E}^2_{\mathcal{L}}}{\textbf{B}^2_{\mathcal{L}}}\, ,
\eeq

\beq
 \vg = \frac{2G_{\mathcal{L}}}{1+G_{\mathcal{L}}}\, ,
\eeq
where,
\beq
G_{\mathcal{L}} = \frac{\Pi^2_{\mathcal{L}}}{(\textbf{D}\phi)^2_{\mathcal{L}}}\, .
\eeq

The subscript $\mathcal{L}$ denotes a Lagrangian density weighting of each magnitude. This is used to ensure that only regions with non-vanishing Lagrangian density contribute to velocities, \ie only strings, where the Lagrangian density is peaked, contribute to the calculation of the estimators. For a given quantity $A$ the weighting is applied in the following way \cite{Hindmarsh:2008dw}:

\beq
A_{\mathcal{L}} = \frac{\int d^3 x A \mathcal{L}}{\int d^3 x \mathcal{L}}\, .
\label{eq_law}
\eeq

All values obtained are shown in Table~\ref{table_velocities}. Fig.~\ref{fig_v} also shows the evolution of the two velocity estimators in the matter domination era, field based in green and gradient based in red, for combined simulations (left panel) and normal simulations (right panel). The curves of the velocities are nearly flat for the whole core-growth phase, \ie natural evolution of the equations of motion. Furthermore, for both estimators the curves are very similar, pointing to a mean value of $\vf\sim\vg\sim 0.26$ (see Table~\ref{table_velocities}). There is no difference in the velocity distribution between combined cases and normal simulations. 

\begin{table*}[h!]
\centering
\scalebox{0.70}{
\renewcommand{\arraystretch}{1.3}
\begin{tabular}{|c|c|c|c|c|c|c|c|c|}
\hline
  & \multicolumn{4}{c|}{Matter} & \multicolumn{4}{c|}{Radiation} \\ \hline
  & \multicolumn{2}{c|}{Combined} & \multicolumn{2}{c|}{Normal} & \multicolumn{2}{c|}{Combined} & \multicolumn{2}{c|}{Normal} \\ \hline
  & $\kappa=0.9$ & $\kappa=0.95$  & $\kappa=0.9$ & $\kappa=0.95$ & $\kappa=0.9$ & $\kappa=0.95$ & $\kappa=0.9$ & $\kappa=0.95$ \\ \hline
$\vf$ & 0.259$\pm$0.005 & 0.261$\pm$0.006 & 0.261$\pm$0.007 & 0.264$\pm$0.006 & 0.306$\pm$0.004 & 0.307$\pm$0.006 & 0.303$\pm$0.004 & 0.305$\pm$0.006\\ \hline  
$\vg$ & 0.259$\pm$0.005 & 0.260$\pm$0.006 & 0.264$\pm$0.007 & 0.266$\pm$0.006 & 0.307$\pm$0.004 & 0.307$\pm$0.005 & 0.306$\pm$0.004 & 0.310$\pm$0.006\\ \hline 
$\vpqf$ & 0.28-0.37 & 0.29-0.35 & 0.28-0.35 & 0.28-0.34 & 0.36-0.44 & 0.36-0.41 & 0.35-0.41 & 0.35-0.40 \\ \hline 
$\vpqg$ & 0.27-0.34 & 0.27-0.32  & 0.28-0.33 & 0.27-0.31 & 0.33-0.39 & 0.33-0.38 & 0.33-0.37 & 0.33-0.37 \\ \hline 
\end{tabular}}
 \caption{\label{table_velocities} Mean values of the field $\vf$ and gradient $\vg$ velocity estimators of the whole network and of field $\vpqf$ and gradient $\vpqg$ velocity estimators of $pq$-strings,   for the different possibilities simulated, averaged over 5 realizations. In the case of   velocity estimators for the whole network we also include $1\sigma$ errors obtained by averaging the standard deviations in the range $\tau > 200$. The two values of the $pq$ estimators correspond to the minimum and maximum mean values respectively.}
\end{table*}

\begin{figure}[h!]
\centering
\includegraphics[width=0.49\textwidth]{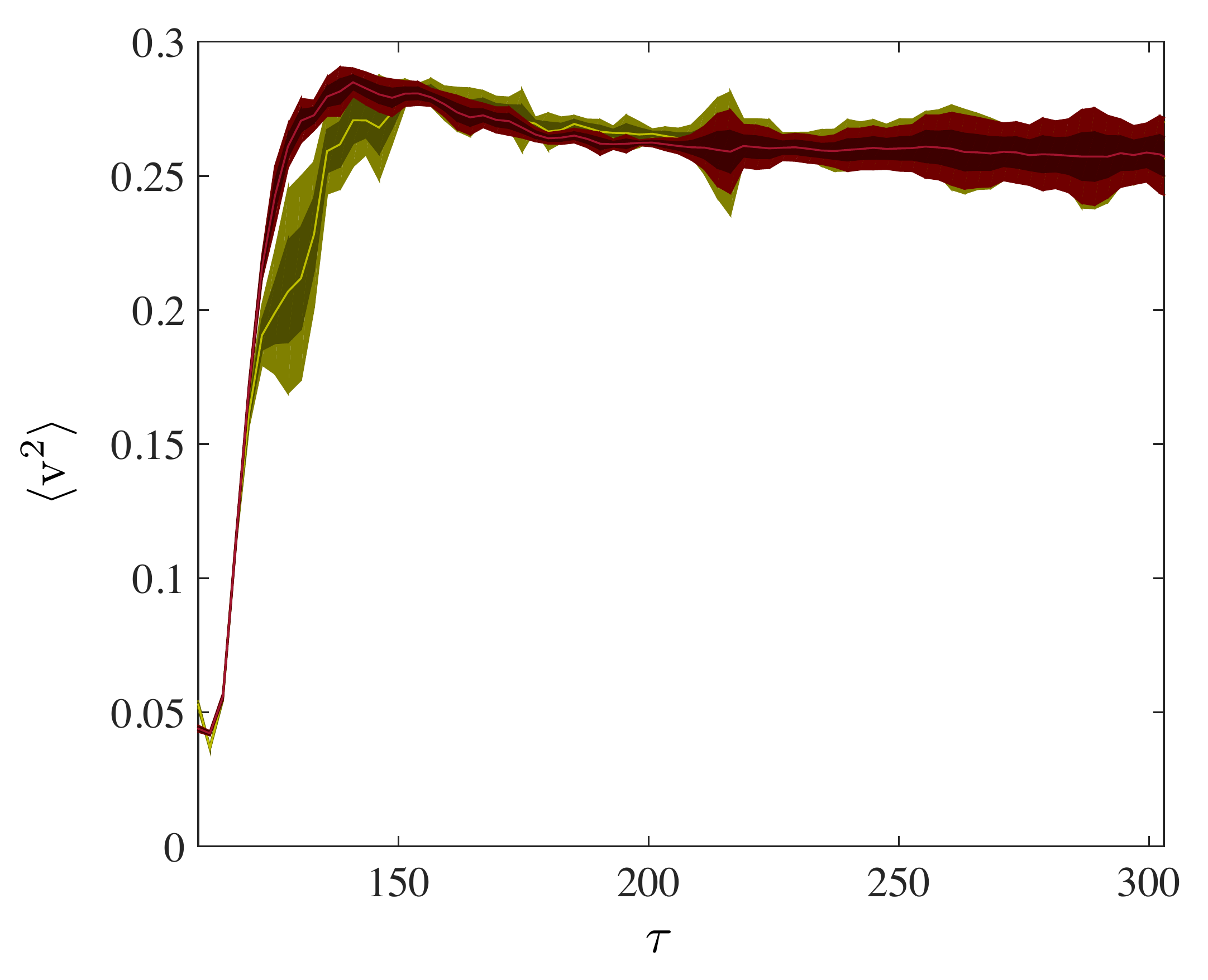}
\includegraphics[width=0.49\textwidth]{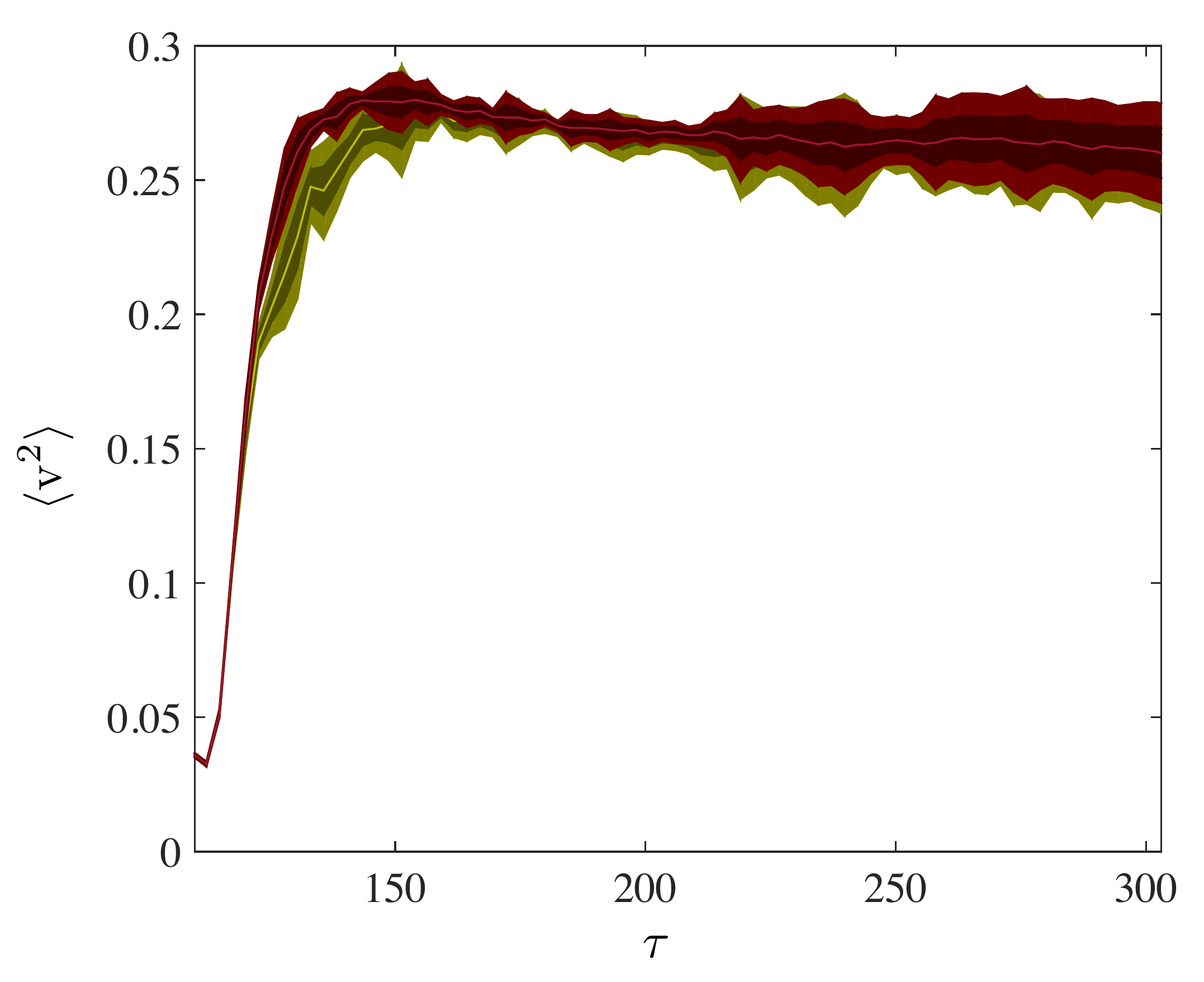}
\caption{Field (green) and gradient (red) Lagrangian weighted velocity estimators of the whole system in matter domination era. The left panel shows the result for combined simulations and the right panel for normal simulations.}
\label{fig_v}
\end{figure}

We also propose a similar estimator for the velocity of the $pq$-strings, $\vpq$. This estimator is based on the previous ones, but the Lagrangian weighting has been substituted by the interaction potential weighting. Hence Eq.~(\ref{eq_law}) is converted into:

\beq
A_{V_{\rm int}} = \frac{\int d^3 x A V_{\rm int}}{\int d^3 x V_{\rm int}}\, .
\label{eq_law1}
\eeq

We have previously observed that $pq$-strings can be well located finding places where the value of the interaction potential is bigger than a threshold value $V_{\rm int}=0.855$ obtained by inspection. Therefore we use this value to focus only on the velocity contribution made by bound states. Nevertheless, the $pq$-string velocity estimator is a mere tentative and as such the values associated to it should not be interpreted as exact estimations.

The results can be found in Table~\ref{table_velocities}. In Fig.~\ref{fig_vpq} we again compare combined simulations (left panel) with normal simulations (right panel), both in matter era.

\begin{figure}[h!]
\centering
\includegraphics[width=0.49\textwidth]{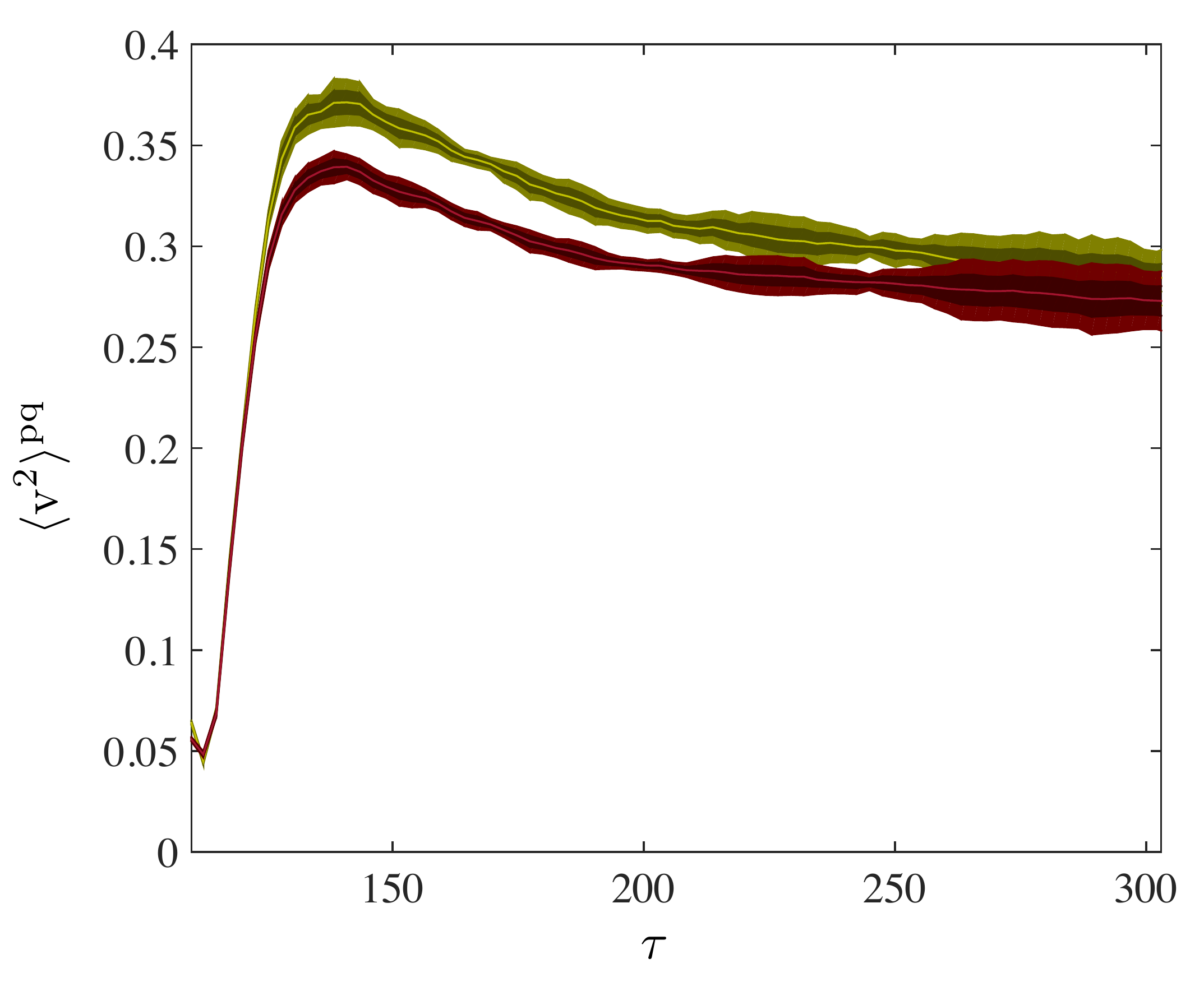}
\includegraphics[width=0.49\textwidth]{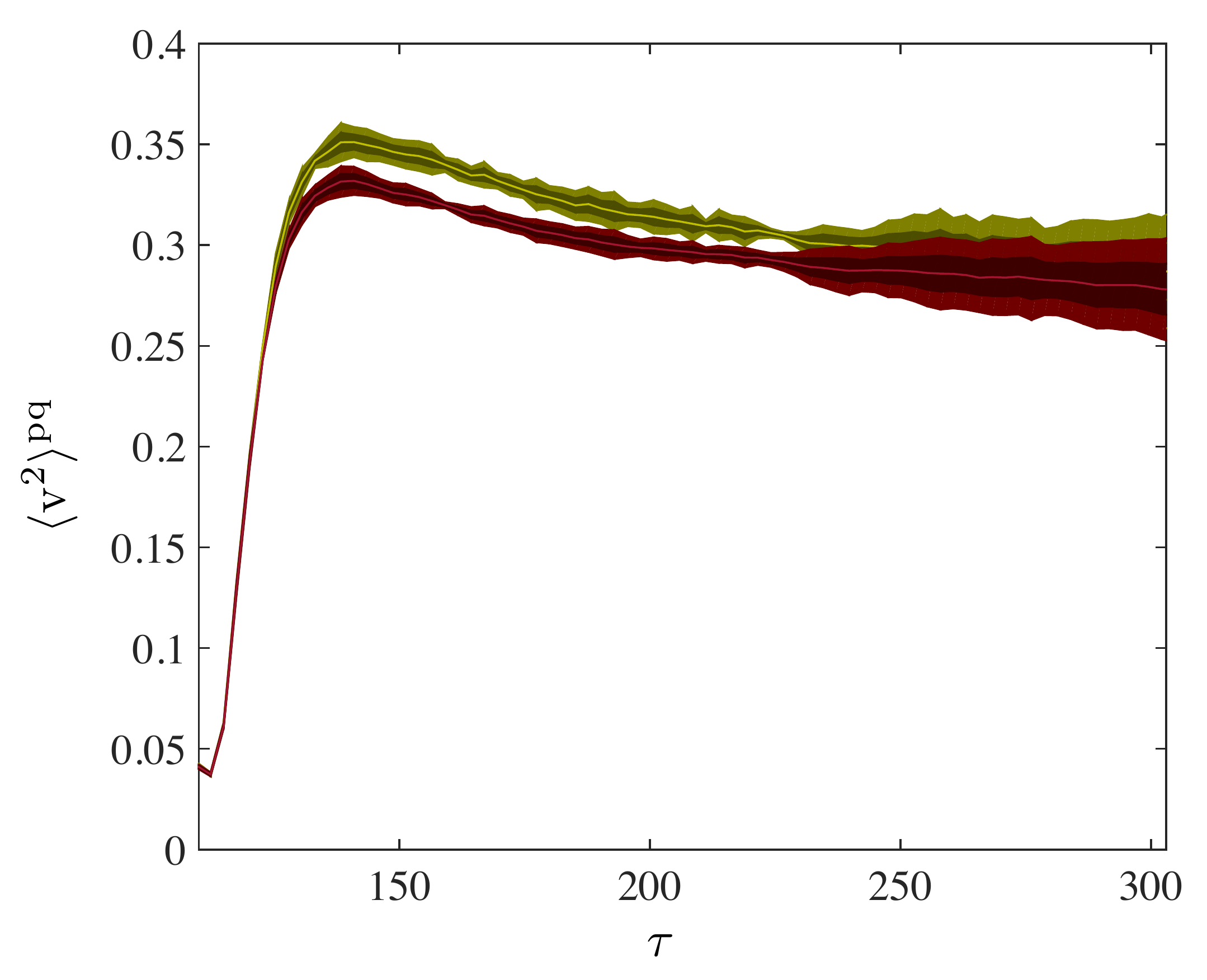}
\caption{Field (green) and gradient (red) velocity estimators weighted by the interaction potential for the $pq$-string network, in matter domination era. Left panel shows the result for combined simulations and right panel for normal simulations.}
\label{fig_vpq}
\end{figure}

In contrast to the estimators of the whole system, velocity estimators of the $pq$-strings show a decreasing tendency. Nevertheless their values lie always above those of the whole system, as Table.~\ref{table_velocities} shows. This decreasing trend can be understood taking into account that the interaction potential weighting also accounts for contributions of crossings of individual $p$- and $q$-strings, this is why their value is higher initially (more crossings, $\mathcal{O}(10^3)$) and lower in the final stages of the simulation ($\mathcal{O}(10^2)$), where the contribution is mainly due to the velocity of the segments. The difference between $\vpqf$ and $\vpqg$ might be caused by the uncertainty of the interaction term weighting approximation. 

Similarly to the case of estimators for the whole system, we observe little differences between combined simulations and normal simulations. The only difference is that the maximum peak in the velocity is slightly higher in the combined case, which might be a consequence of significantly higher density of $pq$-segments at the initial phases of the combined scenario.

As it can be seen in Table~\ref{table_velocities}, simulations in radiation dominated cosmologies show $10\%-15\%$ higher velocities, which have also the same evolution pattern of the matter dominated simulations. We also computed velocities for $\kappa=0.95$, showing no differences with those obtained for $\kappa=0.9$, all within $1\sigma$.


\section{Discussion}
\label{sec_discussion}

In this work we have investigated the survival of bound states in field theoretical models of interconnected $FD$-string networks. These networks  are predicted in several brane inflation models where stable bound states can form as a consequence of the joining of individual $p$ $F$-strings and $q$ $D$-strings and are called $pq$-strings. The properties of these objects are not totally understood, and one of the most interesting methods to explore some of these properties  (but by no means all) are by field theoretical simulations. 

There have been several previous attempts at simulating the properties of $FD$ string networks \cite{ Hindmarsh:2006qn,Rajantie:2007hp,Urrestilla:2007yw,Sakellariadou:2008ay} that confirmed the formation of $pq$-strings and observed that the evolution of the network is compatible with the scaling solution. 
Those works also found that, on the one hand, the relative abundance and length of $pq$-strings were rather low, which questioned the relative importance of the bound string in the evolution of the whole system. On the other hand they showed that although bound strings formed continuously due to the interaction potential, their lifetime was small, indicating the existence of an efficient unzipping mechanism. 

The simulations in \cite{Urrestilla:2007yw} started with a configuration where a network of $p$- and $q$-strings would evolve to form bound states. In this work, we analyze the problem from an alternative (or even opposite) point of view. We explore a scenario where $pq$-strings are present nearly from the beginning of the simulations, instead of waiting for their formation from individual $p$- and $q$-strings. We call this procedure string combination, where one of the individual string networks is replaced by a whole network of bound states interacting with the remaining individual network. We have performed combined simulations in matter and radiation domination eras and compared them with simulations where strings were not combined. 

The objectives we seek with this procedure are twofold: first to see if the low relevance of the bound states is just a consequence of the initial conditions or on the contrary it is product of the natural evolution of this model. Second to determine the efficiency of the unzipping of the composed bound states into individual vortexes, comparing the initial and final bound string abundances.

The main conclusion that can be extracted from combined simulations is that the system is not able to retain a high number of $pq$-strings and tends to split the composed bound states into its constituents. Even though one of the two string networks of the system initially is completely at bound states, the $pq$-string fraction soon starts  to decrease drastically. It is evident that the system does not feel {\it comfortable} with that configuration, and prefers to break the composed bound states into individual $p$- and $q$-strings. By the end of the simulation the relative length on bound states with respect to the total length of the system falls, only $5\%$ of the total length is at $pq$-strings.

The splitting of the $pq$-strings evidences that the system has a great ability to unzip the strings, {\it i.e.}, in our model, the  dynamics of the  Y-junctions at the boundaries of the $pq$-strings is more influenced by the tension applied by the individual free $p$- and $q$-strings than by the dynamics of the heavy string. The binding energy of the bound strings does not seem  high enough for the system to prefer continuing to have those bound states, the force that the constituents of the bound states apply to the bound states, force due to the dynamics of the strings in the network, is more effective in unzipping strings than the effect of the binding energy. Therefore we believe that an efficient unzipping mechanism should also be included in any reasonable effective model that aims to describe interconnected string networks.

Combined simulations  did also exhibit scale invariant evolution. The path towards the scaling regime, though, is somewhat different of that followed by normal simulations. Soon after strings have been combined, the system passes through a relaxation period where most of the $pq$-segments disappear and are converted into ordinary $p$- and $q$-strings. We observe several hints that point to this relaxation period such as the evolution of the fractional amount of $pq$-strings, the characteristic length of $p$-strings or the scaling of $pq$-segments. After the system relaxes the scaling regime is achieved. Remarkably, not only the whole system evolve in a scale invariant manner, but scaling is also achieved by every separate sub-network of the system, especially by the $pq$-string network. In this last case, the characteristic length of the $pq$ network evolves in an almost perfectly linear way, whereas for the normal simulation it can be barely approximated to a straight line (see Fig.~\ref{fig_pqscaling}). 

Another interesting outcome of combined simulations is that the average physical length of the $pq$-strings tends to an asymptotic value of $\lpq\sim55$ in matter domination and $\lpq\sim45$ in radiation domination. Its evolution is also clearly compatible with the relaxation period mentioned in previous lines. Conversely, the evolution of this quantity in normal simulations is clearly an increasing function of time, nearly linear. However, it can be observed that its value lies always below the asymptotic value obtained in the combined case.

The  innovations introduced by combined simulations are very useful to optimize the dynamical range/computational cost of such numerical experiment, as shown by the better scaled achieved by both the network as a whole, and the sub-networks.  The scaling of normal simulations is reached later,  and we postulate that in fact, the attractor solution has not been reached in full at the end of those simulations. In fact, quantities to measure the  characteristic lengths of the $pq$-bound segments  (like $\lpq$  and $\ftot$) disagree at the end of both types of simulations, though they seem to tend towards the same asymptotes, and one could speculate that larger dynamical ranges (specially in normal simulations) could show how these quantities do actually agree in the scaling regime.

Velocities of interconnected string networks and $pq$ bound states have also been analyzed for the first time. We measure the root mean square velocities $\sim 0.5$ for the average velocity of the whole system and depending on the estimator between $\sim 0.5 - 0.6$ for the velocities of the bound states.  Whilst the curves of the average velocities of the system are nearly flat, the velocities of the bound states exhibit a decreasing trend. However, the estimator for the $pq$-string network is not robust, and the velocities obtained should only be considered as a first tentative estimation. Comparisons of the combined and normal simulations show no differences, which enforces our opinion of the validity of string combination method to describe interconnected networks.

The fact that the velocity of $pq$-strings seems higher than that of single $p$- and $q$-strings deserves further investigation. This could go in two directions: on the one hand having larger simulations would allow us to study them for larger periods of time and have a better handle on the estimators. On the other, the dependency of the  velocity of the bound string on its mass (or binding energy) could be crucial to understand its scaling mechanism. It is interesting to study whether the energy liberated in forming the bound states goes into the velocity of the bound state, or it is radiated away similarly to the single strings, or there is yet another mechanism taking part in the process. 

Since the binding energy is indeed a key quantity in our analysis, we tried to increase it within this model by simulating cases for a larger value of the interaction potential coupling constant. We found that increasing the value from $\kappa=0.9$ to $\kappa=0.95$, which is close to the allowed maximum value, does not produce significant changes neither in the amount and lifetime nor in the relative relevance of the $pq$-strings in the global network evolution.  Larger bound energies are necessary to see a significant change in the dynamics of bound states. As it was pointed in \cite{Saffin:2005cs}, the formation of stable $pq$-strings is more favorable for strings with higher winding number than unity. However, no simulation performed during this work produced strings with winding number greater than one. These facts evidence that even though the model  is able to produce $pq$-strings through its interaction potential, for the values of the model chosen in this work the binding energy of $pq$-strings is not high enough to render a high population of them.
 
Nevertheless, it is worthwhile to point that this model is not only limited to the parameter values employed during this work. On the contrary, it can be easily extended to different scenarios which could possibly increase the binding energy of bound states. For instance, simulations with different coupling constants would lead to the departure from the Bogomol'nyi limit. It would be interesting to investigate whether simulations out of the Bogomol'nyi bound are able to produce strings with higher winding numbers. An alternative set of models can be obtained by variation of the vacuum expectation values of the complex scalar fields. Systems composed by scalar fields with unequal vacuum expectation values would produce interconnected string networks composed of {\it individual} strings with different tensions, which would lead to the formation of heavier bound states. Moreover, networks of strings with different masses are expected to be closer to realistic superstring networks. It would be desirable that future field theoretical simulations exploit the limits of the parameter space of the model in order to produce more stable bound states that would play a relevant role in the dynamics of the system. 

\section*{Acknowledgements}
We thank Asier Lopez-Eiguren for useful discussion. This work
has been possible thanks to the computing infrastructure
of the i2Basque academic network and the COSMOS
Consortium supercomputer (within the DiRAC Facility
jointly funded by STFC and the Large Facilities Capital
Fund of BIS). The authors acknowledge support
from the Basque Government (IT-559-10),  the Spanish
Ministry (FPA2012-34456) and  UPV/EHU UFI 11/55.

\bibliographystyle{h-physrev4} 

\bibliography{PQBiblio}

\end{document}